\documentstyle[prd,aps,eqsecnum,epsf,psfig]{revtex}

\newcommand{\bl}{{\bbox{\lambda}}}

\begin{document}
\tighten

\date{\today}
\title{Searching for periodic sources with LIGO}
\author{Patrick R Brady${}^{(1)}$,
	Teviet Creighton${}^{(1)}$,
	Curt Cutler${}^{(2),(3)}$
	and Bernard F Schutz${}^{(3)}$}
\address{${}^{(1)}$ Theoretical Astrophysics 130-33,
	California Institute of Technology,  Pasadena, CA 91125\\
	${}^{(2)}$ Center for Gravitational Physics and Geometry,
	Pennsylvania State University, University Park, PA 16802 \\
	${}^{(3)}$ Max Planck Institute for Gravitational Physics,
	Albert Einstein Institute, Schlaatzweg 1, D-14473 Potsdam
}

\maketitle

\widetext
\begin{abstract}
\hfil
\parbox{5.6in}{We investigate the computational requirements for
all-sky, all-frequency searches for gravitational waves from spinning
neutron stars, using archived data from interferometric gravitational
wave detectors such as LIGO.  These sources are expected to be weak,
so the optimal strategy involves coherent accumulaton of
signal-to-noise using Fourier transforms of long stretches of data
(months to years).  Earth-motion-induced Doppler shifts, and intrinsic
pulsar spindown, will reduce the narrow-band signal-to-noise by
spreading power across many frequency bins; therefore, it is necessary
to correct for these effects before performing the Fourier transform.  The
corrections can be implemented by a parametrized model, in which one
does a search over a discrete set of parameter values ({\em points} in
the parameter space of corrections).  We define a metric on this
parameter space, which can be used to determine the optimal spacing
between points in a search; the metric is used to compute the
number of independent parameter-space points $N_p$ that must be
searched, as a function of observation time $T$.  This method
accounts automatically for correlations between the spindown and
Doppler corrections.  The number $N_p(T)$ depends on the maximum
gravitational wave frequency and the minimum spindown age
$\tau=f/\dot{f}$ that the search can detect.  The signal-to-noise
ratio required, in order to have $99\%$ confidence of a detection,
also depends on $N_p(T)$.  We find that for an all-sky, all-frequency
search lasting $T=10^7$ s, this detection threshhold is $h_c \approx
(4--5) h_{\rm \scriptsize 3/yr}$, where $h_{\rm \scriptsize 3/yr}$ is the
corresponding $99\%$
confidence threshhold if one knows in advance the pulsar position and
spin period.

We define a coherent search, over some data stream of length $T$, to be
one where we apply a correction, followed by an FFT of the data, for
every independent point in the parameter space.  Given realistic
limits on computing power, and assuming that data analysis proceeds at
the same rate as data acquisition (e.g. $10$ days of data gets
analysed in $\sim 10$ days), we can place limitations on how much data
can be searched coherently.  In an all-sky search for pulsars having
gravity-wave frequencies $f\le 200\mbox{\rm Hz}$ and spindown ages $\tau\ge
1000\mbox{\rm Yrs}$, one can coherently search $\sim 18$ days of data on a
teraflops computer.  In contrast, a teraflops computer can only
perform a $\sim 0.8$-day coherent search for pulsars with frequencies
$f \le 1\mbox{\rm kHz}$ and spindown ages as low as $40\mbox{\rm Yrs}$.

In addition to all-sky searches we consider coherent directed
searches, where one knows in advance the source position but not the
period.  (Nearby supernova remnants and the galactic center are
obvious places to look.)  We show that for such a search, one gains a
factor of $\sim 10$ in observation time over the case of an all-sky
search, given a $1\mbox{\rm Tflops}$ computer.

The enormous computational burden involved in coherent searches
indicates a need for alternative data analysis strategies.  As an
example we briefly discuss the implementation of a simple hierarchical
search in the last section of the paper.  Further work is required to
determine the optimal approach.}\hfil
\end{abstract}

\twocolumn
\narrowtext

\section{Introduction}\label{intro}

The direct observation of gravitational waves is a realistic goal for
the kilometer-scale interferometers which are now under construction
at various sites around the world~\cite{Abramovici_A:1992,%
Bradaschia_C:1990}.  However, the battle to see these waves is not
over when the detectors are constructed and running.  Searching for
gravitational wave signals in the interferometer output presents its
own problems, not least of which is the sheer volume of data involved.

Potential sources of gravitational waves fall roughly into three
classes: bursts, stochastic background, and continuous emitters.

Burst sources produce signals which last for times considerably
shorter than available observation times.  The chirp signals from
compact coalescing binaries belong to this class.  Since theoretical
waveforms, valid during the inspiral phase of the binary evolution,
have been accurately calculated using post-Newtonian
methods~\cite{Blanchet_L:1996}, it is possible to search the data
stream for chirps using matched filtering techniques.  Detailed
studies have been carried out to ascertain the optimal set of search
templates~\cite{Apostolatos_T:1996,Owen_B:1996}, and a preliminary
investigation of search algorithms is now under
way~\cite{Balasubramanian_R:1996}.  Detection of other, not so well
understood, sources in this class---e.g. non-axisymmetric
supernovae---has received limited attention~\cite{Thorne_K:1987}.

Flanagan~\cite{Flanagan_E:1993} has determined how to cross-correllate
the output of two detectors in order to search for a stochastic
background of gravitational radiation, which was implemented by
Compton~\cite{Compton_K:1996} and applied to data taken during a
period of 100 hours by two prototype interferometers detectors in
Glasgow and Garching ~\cite{Nicholson_D:1996}.
In~\cite{Allen_B:1996}, Allen presents a detailed discussion of the
potential significance of detecting a stochastic background.
Compton's work, and simulations performed by Allen, have demonstrated
that this kind of analysis requires minimal computational resources.

In this paper we consider some issues involved in searching for
continuous wave sources.  Throughout our discussion we use pulsars as
a guide to develop a search strategy.

\subsection{Gravitational waves from pulsars}\label{introA}

Rapidly rotating neutron stars (pulsars) tend to be axisymmetric;
however, they must break this symmetry in order to radiate
gravitationally.  The pulsar literature contains several mechanisms
which may lead to deformations of the star, or to precession of its
rotation axis, and hence to gravitational wave emission.  The
characteristic amplitude\footnote{We adopt the definition of $h_c$
provided in Eq.~(50) of Thorne~\cite{Thorne_K:1987}.} of gravitational
waves from pulsars scales as
	\begin{equation}
	h_c \sim \frac{I f^2 \epsilon}{r} \label{eq:0.1}
	\end{equation}
where $I$ is the moment of inertia of the pulsar, $f$ is the
gravitational wave frequency, $\epsilon$ is a measure of the deviation
from axisymmetry and $r$ is the distance to the pulsar.

Pulsars are thought to form in supernova explosions.  The outer layers
of the star crystallize as the newborn pulsar cools by neutrino
emission.  Estimates, based on the expected breaking strain of the
crystal lattice, suggest that anisotropic stresses, which build up as
the pulsar loses rotational energy, could lead to $\epsilon \alt
10^{-5}$; the exact value depends on the breaking strain of the
neutron star crust as well as the neutron star's `geological history',
and could be several orders of magnitude smaller.  Nonetheless, this
upper limit makes pulsars a potentially interesting source for
kilometer scale interferometers. Figure~\ref{figure1} shows some upper
bounds on the amplitude due to these effects.

Large magnetic fields trapped inside the superfluid interior of a
pulsar may also induce deformations of the star.
This mechanism has been explored recently in~\cite{Bonazzola_S:1996},
indicating that the effect is extremely small for standard neutron
star models ($\epsilon\alt 10^{-9}$).

Another plausible mechanism for the emission of gravitational
radiation in very rapidly spinning stars is the
Chandrasekhar-Friedman-Schutz (CFS) instability, which is driven by
gravitational radiation
reaction~\cite{Chandrasekhar_S:1970,Friedman_J:1978}.  It is possible
that newly-formed neutron stars may go through this instability
spontaneously as they cool soon after formation.  The radiation is
emitted at a frequency determined by the frequency of the unstable
normal mode, which may be less than the spin frequency.

Accretion is another way to excite neutron stars into emitting
gravitational waves.  Wagoner~\cite{Wagoner_R:1984} proposed that
accretion may drive the CFS instability.  There is also the
Zimmermann-Szedinits mechanism~\cite{Zimmermann_M:1979} where the
principal axes of the moment of inertia are driven away from the
rotational axes by accretion from a companion star.  Accretion can in
principle produce relatively strong radiation, since the amplitude is
related to the accretion rate rather than to structural effects in the
star.  However, accreting neutron stars will be in binary systems, and
these present problems for detection that go beyond the ones we
discuss in this paper.  We hope to return to the problem of looking
for radiation from orbiting neutron stars in a future publication.

\subsection{Three classes of sources}\label{introB}

Observed pulsars fall roughly into two groups: (i) young, isolated
pulsars having periods of tens or hundreds of milliseconds,  and (ii)
older, millisecond pulsars.  The young pulsars are most likely to
deviate significantly from axisymmetry; however, they are generally
observed to have low frequencies, so that there is a
competition between the frequency, $f$,  and deviation from
axisymmetry, $\epsilon$, in Eq.~(\ref{eq:0.1}). On
the other hand, millisecond pulsars, whose waves are higher in
frequency, tend to be quite old and well annealed into an
axisymmetric configuration.

Radio observations can only probe a small portion of our galaxy in
searching for pulsars.  A significant effect reducing the depth of
radio searches is dispersion of the signal by galactic matter between
potential sources and the earth.  Given current evolutionary scenarios
for pulsars---that they are born in supernova explosions---it seems
likely that most pulsars should be located in the galactic
disk,  and the youngest of these will also be shrouded in a supernova
remnant,  making them invisible to radio astronomers.

\begin{figure}
\psfig{file=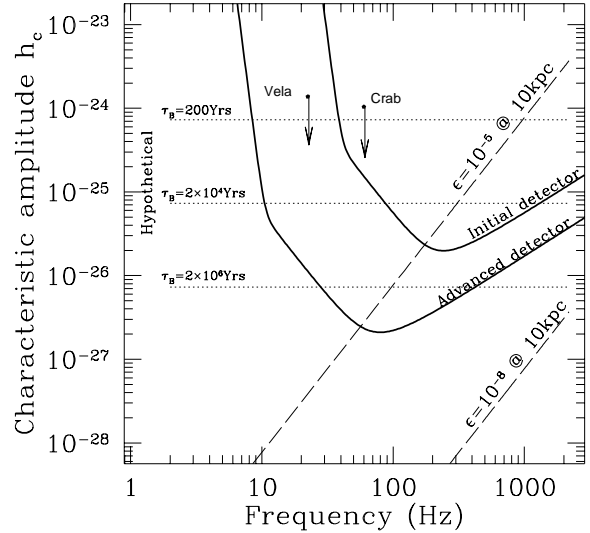,width=8cm,bbllx=18pt,bblly=144pt,bburx=592pt,bbury=718pt}
\caption{\label{figure1} Characteristic amplitudes $h_c$ [see
Eq.~(\protect\ref{eq:3.5a})] for several postulated periodic sources,
compared with sensitivities $h_{\rm \protect\scriptsize 3/yr}$ of the
initial and advanced detectors in LIGO.  ($h_{\rm \protect\scriptsize
3/yr}$ corresponds to the amplitude $h_c$ of the weakest source
detectable with 99\% confidence in $\frac{1}{3} {\rm yr} =
10^7\mbox{\rm s}$ integration time, if the frequency and phase of the
signal is known in advance.)  Long-dashed lines show the expected
signal strength as a function of frequency for pulsars at a distance
of 10~kpc, assuming non-axisymmetries of $\epsilon = 10^{-5}$ and
$\epsilon = 10^{-8}$, where $\epsilon$ is defined in
section~\protect\ref{gravitational-waves}A.  Upper limits are also
plotted for the Crab and Vela pulsars, assuming their entire measured
spindown is due to gravitational wave emission.  The dotted lines
indicate the strongest waves received at the earth for Blandfords
hypothetical class of pulsars; each line corresponds to a particular
birth rate.}
\end{figure}

Blandford~\cite{Blandford_R:1984,Thorne_K:1987} has pointed out that
there could exist a class of pulsars which spin down primarily due to
gravitational radiation reaction.  For sources in this class the
frequency scales as $f\propto\tau^{-1/4}$, where $\tau$ is the age of
the pulsar.  If the mean birth rate for such pulsars in our galaxy is
$\tau_B^{-1}$, the nearest one should be a distance
$r=R\sqrt{\tau_B/\tau}$ from earth, where $R\simeq 10$kpc is the
radius of the galaxy.  The intrinsic gravitational wave amplitude
(that is, the amplitude $h$ at some fixed distance) of a pulsar in
this class is proportional to $\tau^{-1/2}$.  Thus, the nearest
source in this class would have a dimensionless amplitude $h_c$ at the
Earth
	\begin{equation}
	h \simeq 8\times10^{-25} \left( \frac{200 \mbox{\rm Yrs}}{\tau_B}
	\right)^{1/2}
	\end{equation}
independent of the frequency,  or the the ellipticity $\epsilon$
of the pulsar. Assuming the existence of such a
class of pulsars, with
$\tau_B \alt 2\times 10^4$ Yrs, we see from Fig.~\ref{figure1} that there is a
large
region of parameter space that is both (i) detectable by the LIGO
detector and (ii) physically reasonable, in the sense that $\epsilon < 10^{-5}$
and $f$ lies in the range 200--1000~Hz.

 Note that Blandford's argument can be slightly re-cast to
yield an upper limit on the gravitational wave strength of {\it any}
isolated pulsar--i.e., any pulsar whose radiated angular momentum is
not being replenished by accretion. The age of an isolated pulsar must
be shorter than the age computed assuming the spin-down is solely due
to gravitational wave emission.  Correspondingly, if we set $\tau_B$
equal to $40$ yrs (corresponding to the birthrate for $\it all$
pulsars), we get the following upper limit for measured gravitational
wave amplitude of an isolated pulsar: $h_c < 2\times 10^{-24}$.  Of
course, this is a statistical argument.  This bound could certainly be
violated by an isolated pulsar that just happens to be anomalously
close to us.

It is important that any search strategy should be general enough to
encompass all three of the above classes, allowing for the significant
changes in frequency which may be inherent in the sources (see
section~\ref{pulsars}).

\subsection{The data analysis problem}\label{introC}

The detection of continuous, nearly fixed frequency waves will be
achieved by constructing power spectrum estimators and searching for
statistically significant peaks at fixed frequencies.  In practice,
this is achieved by calculating the amplitude of the Fourier transform
of the detector output given by applying a fast Fourier transform
(FFT), a discrete approximation to the true Fourier transform:
	\begin{equation}
	\tilde h(f) = \frac{1}{\sqrt{T}} \int_0^T e^{2\pi i f
	t} \; h(t) \; dt \; . \label{eq:1.1}
	\end{equation}
The main hope of detection lies in the fact that one may observe the
sky for long time periods of time $T$.  When such a data stretch is
transformed to make the underlying signal monochromatic, the signal to
noise ratio grows as $\sqrt{T}$ in amplitude (or as $T$ in the power
spectrum).  One will likely need to have integration times of several
weeks or months in order for the expected signals from nearby sources
to rise above the noise.  However, such long data stretches pose a
significant computational burden; using $10^7\mbox{\rm s}$ of data to
look for signals with gravitational wave frequencies up to
$500\mbox{\rm Hz}$ requires calculating an FFT with $N\simeq 10^{10}$
data points.  Calculation of a single such FFT would take about
$1\mbox{\rm s}$ on a $1$Tflops computer, assuming that all $10^{10}$
points can be held in fast memory.  Unfortunately, this is not the
whole story.

The detection problem is complicated by the fact that the signal
received at the detector is not perfectly monochromatic.  Earth-bound
detectors participate in complex motions which lead to significant
Doppler shifts in frequency as the Earth rotates, and as it orbits
around the sun (this orbit is significantly perturbed by the moon and
the other planets).  The time dependent accelerations broaden the
spectral lines of fixed frequency sources spreading power into many
Fourier bins about the observed frequency.  In order to maintain the
benefit of long observation times, it is therefore necessary to remove
the effects of the detector motion from the data stream. This can
achieved by introducing an inertial (barycentered) time coordinate and
carrying out the FFT with respect to it.  The difficulty of doing this
was first estimated by one of us~\cite{Schutz_B:1991}.  However, we
must also consider the additional complication that the signal may not
be intrinsically monochromatic.  If the signal exhibits intrinsic
frequency drift, or modulation, due to the nature and location of the
source --- as is expected for pulsars which spin down with time ---
these effects can also be removed in the transformation to the new
time coordinate.

Unfortunately, the demodulated time coordinate depends strongly on the
direction from which the signal is expected, and on the intrinsic
frequency evolution one assumes for the source.  Thus, in searching
for sources whose position and timing are not well known in advance
one must apply many different corrections to the data, performing a
new FFT after each correction.  Given the possibility that the
strongest sources of continuous gravitational waves may be
electromagnetically invisible or previously undiscovered, an {\em all
sky, all frequency\/} search for such unknown sources is of
considerable interest.  To obtain some idea of the magnitude of this
task, consider searching the entire sky for signals with (fixed)
frequencies up to $500\mbox{\rm Hz}$ using $10^7\mbox{\rm s}$ worth of data.
Assuming the entire data stream could be held in fast memory on a
machine capable of $1\mbox{\rm Tflops}$, it would take $10^8\mbox{\rm s}$ to
complete
the search.  Introducing intrinsic spindown effects into the search
increases the computational cost, at fixed integration time, by many
orders of magnitude.  This computational cost is the central problem
of searching for unknown pulsars in the output from gravitational wave
detectors and is the focus of this paper.

\subsection{Summary of results}\label{introD}

We parametrise the space of pulsar signals by the position of the
source on the sky $\{\theta,\phi\}$, entering through Doppler shifts
due to the detector's motion, and by spindown parameters $f_k$ which
characterise the intrinsic frequency evolution.  [See Eq.~(\ref{eq:3.6}).]
We constrain the range of possible values of the spindown parameters
using the (spindown) age $\tau=f/\dot{f}$ of the youngest search that
a search can detect, thus $|f_k| \le \tau^{-k}$.  For the
computationally-intensive search over all sky positions and spindown
parameters, it is important to be able to calculate the smallest
number of independent parameter values which must be sampled in order
to cover the entire space of signals.  We have accomplished this by
introducing a distance measure and corresponding metric on the
parameter space.  The analysis is patterned after a similar one
developed by Owen~\cite{Owen_B:1996} for gravitational waves from
inspiralling, compact binaries.  Using our metric one can compute the
volume of parameter space, thus inferring the number of independent
points that must be sampled in order to cover the entire space.  We
define a {\em coherent search} to be one where we perform one
demodulation and FFT of the data for every independent point in the
parameter space. Besides telling us the computational requirements for
a coherent search, the metric approach tells us how to place the
points most efficiently in parameter space, in a similar way to that
discussed by Owen.

We have found it useful to present the results based on several
possible search strategies, which cover different regions of the
parameter space.  Accordingly, we define a pulsar to be {\em old} if
its spindown age $\tau$ is greater than $10^3$ Yrs and {\em young} if
$\tau \agt 40$ Yrs. A pulsar is considered to be {\em slow} if its
gravitational wave frequency is $f \alt 200$ Hz and {\em fast} if $f
\alt 10^3$ Hz.

A coherent all-sky search of $10^7$ seconds of data for old, slow
pulsars requires approximately $1.1\times10^{10}$ independent points
in the parameter space; only one spindown parameter is needed to
account for intrinsic frequency evolution.  In contrast, an all-sky
search for fast, young pulsars in $10^7$ seconds of data requires
$8\times10^{21}$ independent parameter space points to be sampled,
using three spindown parameters to model intrinsic frequency
evolution.  Note that searches for old, fast pulsars (such as known
millisecond radio pulsars) and young, slow pulsars (younger brothers
of the Crab and Vela) are automatically subsumed under the latter
search.  These results mean the following.  Assuming unlimited
computer power and stationary, gaussian statistics, a pulsar with
unknown position and period must have strain $h_c \approx 4.3
h_{\rm \protect\scriptsize 3/yr}$, if it is in our `old, slow' category, and
$h_c \approx 5.1
h_{\rm \protect\scriptsize 3/yr}$, if it is in our `young, fast' category, to
be detected with
$99\%$ confidence in a $10^7$ second search.  Here $h_{\rm \protect\scriptsize
3/yr}$ is the
strain required for detection with $99
\%$-confidence in a $10^7$ second integration, assuming the
pulsar position and period are known in advance\footnote{This differs
from equation (112) in~\cite{Thorne_K:1987} because we have specified
$99\%$ confindence,  and we have use the correct exponential
probability function for power.}:
	\begin{equation}
	h_{\rm \protect\scriptsize 3/yr} (f) = 
	4.2\, \sqrt{S_n(f) \times 10^{-7} Hz} \; .
	\label{eq:1.2}
	\end{equation}
Thus, when considering an all-sky, all-frequency pulsar search, the LIGO
sensitivity curves shown in Fig.1 effectively overestimate the
detector's sensitivity by a factor of $\sim 4--5$, even in the limit of
infinite computing power.

Our ability to perform searches for continuous waves will certainly be
limited by the available computing resources.  Assuming realistic
computer power --- say of order $10^{13}$ flops --- we estimate that
computing limitations will effectively reduce the sensitivity of the
detector by another factor of $\sim 2$, even for some reasonably
optimized and efficient search strategy.  However more work will be
needed to develop a reasonably optimized algorithm, and thus to refine
this latter estimate.

While the concept of the metric is introduced in the framework of an
all-sky search for unknown pulsars, it is clear that we may use the
same approach to examine the depth of a search over limited regions of
the parameter space.  In particular, once the scope of a search is
decided, the optimization procedure dicussed in section~\ref{all-sky}
can be used to determine the observation time and grid spacing which
maximises the expected sensitivity of a search.  As an example, we
consider coherent {\it directed searches}, in which one assumes a
specific sky position (such as a particular cluster or supernova
remnant) and searches only over spindown parameters. Again, we present
results for two concrete scenarios based on fast, young pulsars and
old, slow pulsars.  Similar considerations apply to directed searches
as to all-sky searches; that is, the curves in Fig.~\ref{figure1}
overestimate the detector sensitivity for $10^7$ second integration.
Table~1 summarises the results for both cases.

\begin{figure}
\psfig{file=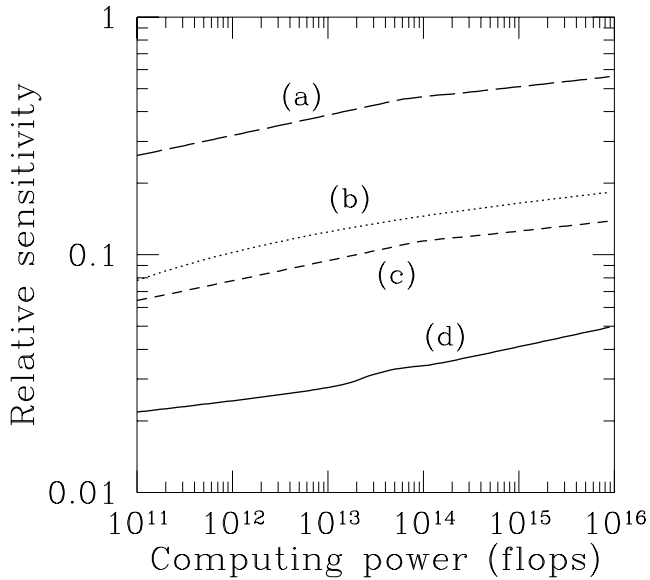,width=8cm,bbllx=18pt,%
bblly=144pt,bburx=472pt,bbury=718pt}
\caption{\label{figure1a} Relative amplitude sensitivities
$h_{\rm \protect\scriptsize 3/yr}/h_{\rm \protect\scriptsize 1/T}$ achievable
with given
computational resources, for various coherent search strategies: (a)~directed
search for old ($\tau\ge 1000\mbox{\rm Yrs}$), slow ($f\le 200\mbox{\rm Hz}$)
pulsars,
(b)~all-sky search for old, slow pulsars, (c)~directed search for
young ($\tau\ge 40\mbox{\rm Yrs}$) fast ($f\le 1000\mbox{\rm Hz}$) pulsars,
and,
(d)~all-sky search for these same sources.  $h_{\rm \protect\scriptsize 1/T}$
is
the expected sensitivity of the detector for an observation time $T$
seconds, and with $99\%$ confidence, assuming only that the frequency
bandwidth of the source is constrained in advance.  For a given
computational power, we have determined the optimum observation time
as described in sections~\protect\ref{all-sky}B and
\protect\ref{directed}. }
\end{figure}

We note that in each type of search, the number of parameter space
points, and hence the computational requirements, were reduced
significantly by the assumption that the points were placed with
optimal spacings given by the metric formalism.  Nevertheless, the
bottom line is that limitations on computational resources will
severely restrict the integration times that can be achieved.
Assuming access to a \mbox{\rm Tflops}\ of computing power (effective
computational throughput, ignoring possible overheads due to
interprocessor communication or data access), we find the following
limits on coherent integration times: For young, fast pulsars we are
limited to about 0.8 days for an all-sky search, and 18 days for a
directed search.  For older, slower pulsars, on the other hand, we are
only limited to 9 days for an all-sky search, and nearly 160 days for
a directed search.  The threshold sensitivities that these strategies
can achieve, relative to the noise curves in Fig.~\ref{figure1}, are
plotted as functions of computing power in Fig.~\ref{figure1a}.

\begin{table}
\caption{\label{table1} The number of independent parameter points
$N_p(T, \mu_{\rm \protect\scriptsize max} = 0.3)$ required for a coherent $T =
10^7 {\rm s}$
search, for four fiducial types of pulsar.  We list the requirements
both for all-sky searches and for directed searches (i.e., searches
where the source position is known in advance). Also listed are the
threshhold values $h_{\rm \protect\scriptsize th}$ of the characteristic strain
$h_c$ required
to have $99\%$ confidence of detection, assuming unlimited computer
power.  These threshold values are given by
$h_{\rm \protect\scriptsize th}/h_{\rm \protect\scriptsize 3/yr} =
(1/1.90)\protect\sqrt{ \ln(50 N N_p) - 1}$
where $N \equiv 2 f_{\rm \protect\scriptsize max} T$.
Here $h_{\rm \protect\scriptsize 3/yr}$ is the corresponding threshhold,
assuming the
pulsar's postion and period and are known in advance.
}
\begin{tabular}{cccccc}
\tableline
\multicolumn{2}{c}{Search Parameters} & $N_p$ & $h_{\rm \protect\scriptsize
th}/h_{\rm \protect\scriptsize 3/yr}$ &
$N_p$ & $h_{\rm \protect\scriptsize th}/h_{\rm \protect\scriptsize 3/yr}$ \\
f (Hz) & $\tau$ (Yrs) &(all-sky)&(all-sky)&(directed)&(directed)\\
\tableline
$< 200$  & $> 10^3$ & $1.1 \times 10^{10}$ & $3.7$
                    & $3.7 \times 10^6$    & $3.3$ \\
$< 10^3$ & $> 10^3$ & $1.3 \times 10^{16}$ & $4.2$
                    & $1.2 \times 10^8$    & $3.5$ \\
$< 200$  & $> 40$   & $1.7 \times 10^{18}$ & $4.3$
                    & $8.5 \times 10^{12}$ & $3.9$ \\
$< 10^3$ & $> 40$   & $8 \times 10^{21}$   & $4.6$
                    & $1.4 \times 10^{15}$ & $4.1$ \\
\tableline
\end{tabular}
\end{table}

\subsection{Organization of this paper}\label{introE}

In section~\ref{pulsars} we outline the physics of pulsars which is relevant to
the detection of continuous gravitational waves.  The discussion is
phenomenological and based almost entirely on pulsar data collected by
radio astronomers.  We focus attention on effects which may lead to
significant frequency evolution over periods of several weeks of
observation.

Then, in section~\ref{gravitational-waves}, we introduce a
parameterised model of the expected gravitational waveform, including
modulating effects due to detector motion.

{}From this, we go on in section~\ref{data-analysis-technique} to describe the
basic technique used to search for signals, by constructing a
demodulated time series.  Livas~\cite{Livas_J:1987},
Jones~\cite{Jones_G:1995} and Niebauer~\cite{Niebauer_T:1993} have
implemented variants of this basic search strategy over limited
regions of parameter space (in particular they have not considered
pulsar spin-down, and have restricted attention to small areas of the
sky).

For the more computationally-intensive search over all sky positions
and spindown parameters, it is important to be able to calculate the
smallest number of independent parameter values which must be sampled
in order to cover the entire space of signals.  In section~\ref{metric} we
develop the metric formalism for calculating the number of independent
points in parameter space.

In sections~\ref{all-sky} and \ref{directed} we apply this formalism
to determine the computational requirements of an all-sky search for
unknown pulsars and a directed search, respectively.

Finally in section~\ref{future}, we list some possible alternatives to a
straightforward coherent search of the interferometer data.  Detailed
studies of the pros and cons of each are currently under
investigation.

\section{Pulsar phenomenology}\label{pulsars}

Currently, the only expected sources of continuous, periodic
gravitational waves in the LIGO band are pulsars.  In this section,
therefore, we review those properties of pulsars which may be
important in the detection process.  In general, the search technique
we present later is capable of detecting any {\em nearly\/}
monochromatic gravitational wave with sufficient amplitude. However,
it is useful to have a concrete physical system in mind when
considering the expected gravitational waveform.

That pulsars are rapidly rotating neutron stars is now well
established~\cite{Shapiro_SL:1983}.  Their high densities and strong
gravitational fields allow them to withstand rotation rates of
hundreds of times per second.  Moreover, pulsar emission mechanisms
require large magnetic fields, frozen into (co-rotating with) the
neutron star.  Indeed these large field strengths may produce
non-axisymmetric deformations of the pulsar.  However, the most
remarkable feature of pulsars is the very precise periodicity of
observed pulses.

There are more than 700 known pulsars, all at galactic distances,
concentrated in the galactic plane.  Based on the sensitivity limits
of radio observations the total number of active pulsars in our galaxy
is estimated to be more than $10^5$~\cite{Lyne_A:1992,Narayan_R:1990}.

\subsection{Spindown}\label{spindowns}

\begin{figure}
\psfig{file=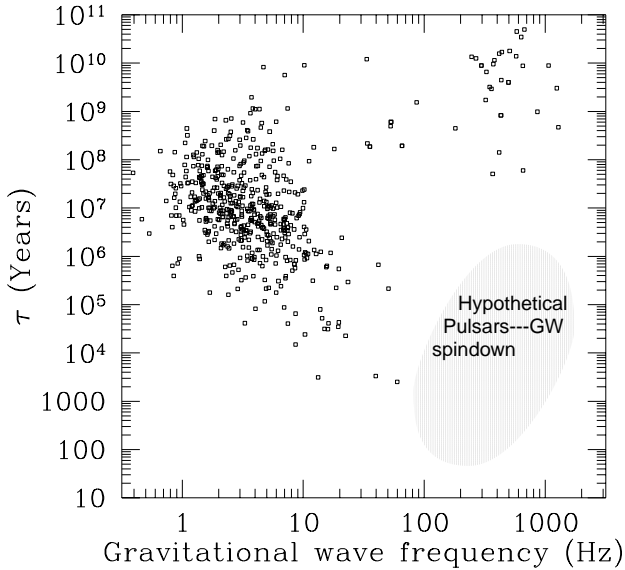,width=9cm,bbllx=0in,bblly=1.5in,bburx=8.5in,bbury=9.5in}
\caption{\label{figure2}
Gravitational wave frequency versus spindown age, $\tau=f/(df/dt)$,
measured in years, for 540 pulsars which have measured period
derivative.  The figure clearly shows a large concentration of pulsars
in the mid-left of diagram.  Most of these are isolated pulsars.  The
standard evolutionary scenario suggests that pulsars move from higher
frequencies and shorter spindowns left and up towards this main bunch.
In contrast, many of the millisecond pulsars lying in the upper right
of the figure are in binary systems, and it is widely believed that
these are pulsars which have been spun up by mass accretion from the
companion star.}
\end{figure}

Pulsars lose rotational energy by electromagnetic braking, the
emission of particles and, of course, emission of gravitational
waves~\cite{Manchester_R:1992,Kulkarni_S:1992}.  Thus, the rotational
frequency is not completely stable, but varies over a timescale $\tau$
which is of order the age of the pulsar.  Typically, younger pulsars
(with periods of tens of milliseconds) have the largest spindown
rates.  Figure~\ref{figure2} shows the distribution of rotational
frequencies and spindown age, $\tau=f/(df/dt)$.

Current observations suggest that spindown is primarily due to
electromagnetic braking; however, for detection purposes it is
necessary to construct a sufficiently general model of the frequency
evolution to cover all possibilities.  For observing times
$t_{\rm \protect\scriptsize obs}$, say, much less than $\tau$, the frequency
drift is small
and the rotational frequency\footnote{We choose to parameterise the
frequency by what will be the gravitational wave frequency, $f_0$,
thus introducing the extra factor of $2$ into this expression} can be
modeled as a power series of the form
	\begin{equation}
	f(t) = (f_0/2) \bigl( 1 + \sum_k f_k
	t^k \bigr) \; . \label{eq:2.1}
	\end{equation}
If $\tau_{\rm \protect\scriptsize min}$ is the shortest timescale over which
the frequency
is expected to change by a factor of order unity, the coefficients satisfy
	\begin{equation}
	| f_k | \alt \tau_{\rm \protect\scriptsize min}^{-k} \; .
	\label{eq:2.2}
	\end{equation}
Clearly,  for an observation time $t_{\rm \protect\scriptsize obs} \ll
\tau_{\rm \protect\scriptsize min}$,
the first few terms in this series will dominate.

Observations suggest that pulsars are born in supernova explosions
with very short periods (perhaps several milliseconds), and
subsequently spin down on timescales comparable to their age.
Supernovae are observed in galaxies similar to our own at the rate of
two or three per century, so we might expect $\tau_{\rm \protect\scriptsize
min} \sim
40$~years for pulsars in our galaxy.  It is at this point that the
distinction between various classes of pulsars becomes important.  The
known millisecond pulsars are old neutron stars which have have been
spun up to periods of only a few milliseconds, possibly by
episodes of mass transfer from a companion star.  As seen from
Fig.~\ref{figure2}, timing measurements of millisecond pulsars yield
very long spindown timescales, $\tau_{\rm \protect\scriptsize min} \agt 10^7$
years.

\subsection{Proper Motions}\label{proper-motions}

Pulsars are generally high velocity objects~\cite{Manchester_R:1992},
as can be inferred by the distance they move in their lifetimes.
Proper motions cause Doppler shifts in the observed pulsar frequency.
If the motion is uniform (constant velocity), it simply induces a
constant frequency shift --- an effect which is undetectable.
However, acceleration and higher order derivatives of the source's motion
will modulate the observed frequency.

Studies of millisecond pulsars in globular clusters have shown that
acceleration in the cluster field can produce frequency drifts which
are comparable in magnitude to the spindown
effects~\cite{Wolszczan_A:1989,Anderson_S:1993}.  Once again, we
expect these effects to be well modelled by a power series in
$t/\tau_{\rm \protect\scriptsize cross}$, where $\tau_{\rm \protect\scriptsize
cross}$ is the time it takes the
pulsar to cross the cluster.  We expect that
$\tau_{\rm \protect\scriptsize min}\le\tau_{\rm \protect\scriptsize cross}$ for
these objects (since if not,
the pulsar will already have escaped the cluster).  Thus the
frequecy model adopted above should be sufficiently general to
encompass the observational effects of proper motions of the sources.

A large proportion of millisecond pulsars are also in binary systems.
Unfortunately, such pulsars participate in proper motions which vary
over very short timescales (their orbital periods).  The
time-dependent Doppler effect due to these motions is {\it not\/}
modelled accurately by a simple power series as in Eq.~(\ref{eq:2.1}).  They
would require a more elaborate model involving as many as five unknown
orbital parameters.  Including these effects in a coherent, all-sky
search strategy would be prohibitive (see section~\ref{all-sky}).  In a search
for gravitational waves from a known binary pulsar, however, it would
be important to deal with this effect.

Proper motions can also affect a search if the star moves across more
than one resolution element on the sky during an observation.  For the
lengths of observation periods envisioned here, this is unlikely to be
a problem.  In an observation lasting a year, however, a pulsar with a
spatial velocity of $1\times10^3\;\rm km\; s^{-1}$ at a distance of
$300\;pc$ will move by about half an arcsecond, which is comparable to
the resolution limit for our observations if the pulsar frequency is
1\mbox{\rm kHz}.

\subsection{Glitches}\label{glitches}

In addition to gradual frequency drifts due to spindown, some young
pulsars exhibit occasional, abrupt increases in frequency.  The
physical mechanism behind these frequency {\it glitches} is not well
understood, although the number of observations of glitch events is
growing~\cite{Lyne_A:1992}.  Given the stochastic nature of glitching,
and the expectation that several months will elapse between major
events, we will ignore glitching in this paper.

\section{Gravitational waves from pulsars}\label{gravitational-waves}

In order to gain insight into the detection problem it is also
important to understand the expected gravitational wave signal.
Several mechanisms have been discussed in the literature which may
produce non-axisymmetric deformations of a pulsar, and hence lead to
gravitational wave
generation~\cite{Bonazzola_S:1996,Chandrasekhar_S:1970,%
Friedman_J:1978,Zimmermann_M:1979,Zimmermann_M:1978,Zimmermann_M:1980}.

In general, a pulsar can radiate strongly at frequencies other than
twice the rotation frequency.  For example, a pulsar deformed by
internal magnetic stresses, which are not aligned with a principal
axis, can radiate at the rotation frequency and twice that
frequency~\cite{Galtsov_D:1984}.  If the star precesses, it will
radiate at three frequencies: the rotation frequency, and the rotation
frequency plus and minus the precession
frequency~\cite{Zimmermann_M:1979}. The important point, however, is
that the signal at the detector is generally narrow band, exhibiting
only slow frequency drift on observational timescales.

Therefore, in this section we outline the main features of the
expected waveform and the corresponding strain measured at a detector
for the case of crustal deformation; other scenarios give similar
results except for the presence of more than one spectral component.

\subsection{Waveform}\label{waveform}

Adopting a simple model of a distorted pulsar as a tri-axial
ellipsoid, rotating about a principal axis with a frequency given
by Eq.~(\ref{eq:2.1}), one may compute the expected gravitational wave signal
using the quadrupole formula.  The two polarizations are
	\begin{eqnarray}
	h_+ &=& h_0 (1+\cos^2 i)\; \cos \{ 2\pi\,f_0\, [ t +
	{\textstyle \sum} f_k{\textstyle\frac{t^{k+1}}{k+1}}
		]\} \; , \label{eq:3.1} \\
	h_\times &=& 2 h_0 \cos i \; \sin \{ 2\pi\,f_0\, [ t +
	{\textstyle \sum} {f_k}
		{\textstyle\frac{t^{k+1}}{k+1}} ]\} \; , \label{eq:3.2}
	\end{eqnarray}
where $i$ is the angle between the rotation axis and
the line of sight to the source.   The dimensionless amplitude is
	\begin{equation}
	h_0 = \frac{8\pi^2 G}{c^4} \; \frac{I_{zz} f_0^2}{r}
		\; \epsilon \; ,\label{eq:3.3}
	\end{equation}
where
	\begin{equation}
	\epsilon = \frac{I_{xx}-I_{yy}}{I_{zz}} \label{eq:3.4}
	\end{equation}
is the {\em gravitational ellipticity} of the pulsar.  The distance to
the source is $r$, and $I_{jk}$ is its moment of inertia tensor.

The strength of potential sources is best discussed in terms of the
characteristic amplitude $h_c$, defined in Eq.~(50)
of~\cite{Thorne_K:1987},  and simply related to $h_0$ by
	\begin{equation}
	h_c = \sqrt{\frac{2}{15}} h_0 \; . \label{eq:3.5a}
	\end{equation}
For a typical $1.4M_{\odot}$ neutron star,
having a radius of $10$km and at a distance of $10$kpc, the
dimensionless amplitude is
	\begin{equation}
	h_c = 7.7\times 10^{-25}\;
	\frac{\epsilon}{10^{-5}}\; \frac{I_{zz}}{10^{45} \mbox{ g
	cm}^2}\frac{10\mbox{kpc}}{r} \left(\frac{f_0}{1\mbox{kHz}}
	\right)^2 \; .  \label{eq:3.5}
	\end{equation}
The magnitude of the gravitational ellipticity, $\epsilon$, represents
the central uncertainty in any estimate of gravitaional waves from
pulsars.  The tightest theoretical constraint, $\epsilon < 10^{-5}$,
is set by the maximum strain that the neutron star crust may
support~\cite{Thorne_K:1987}.  It has also been suggested that
stresses induced by large magnetic fields might result in significant
gravitational ellipticity.  Recently, Bonazzola and
Gourgoulhon~\cite{Bonazzola_S:1996} have considered this possibility,
finding discouraging results; their calculations indicate $10^{-13}\alt
\epsilon \alt 10^{-9}$ depending on the precise model they consider.

In any case, an upper bound on the gravitational ellipticity is
$\epsilon \sim 10^{-5}$, although typical values may be
significantly smaller.

\subsection{Signal at the detector}\label{signal}

Observing the gravitational waves using an earth-based interferometer
introduces two further difficulties into the detection process:
Doppler modulation of the observed gravitational wave frequency, and
amplitude modulation due to the changing orientation of the detector.

For the purpose of detection, the Doppler modulation of the observed
gravitational wave frequency, due to motion of the detector with
respect to the solar system barycenter, is a large effect.  Assuming
the intrinsic frequency model (\ref{eq:2.1}) for the pulsar rotation, the
gravitational wave frequency measured at the detector is
	\begin{equation}
	f_{\rm \protect\scriptsize gw}(t) = 
		f_0 \, \bigl(1 + \sum_k f_k t^k\bigr)
		\bigl(1 + \frac{\vec v}{c}\cdot\hat n\bigr)
		\label{eq:3.6}
	\end{equation}
where $\vec v (t)$ is the detector velocity and $\hat n$ is the unit
vector pointing to the pulsar, in some inertial frame.  We generally
choose this frame to be initially comoving with the Earth at $t=0$.
The frequency measured in this frame is identical to that measured at
the solar system barycenter except for an unimportant constant shift
in $f_0$.

To understand the amplitude modulation we must introduce the Euler
angles, $\{\Theta, \Phi, \Psi\}$, which specify the orientation of the
gravitational wave frame with respect to the detector frame.  The
dimensionless strain at the detector is
	\begin{equation}
	h = F_+(\Theta,\Phi,\Psi) h_+ +
		F_\times(\Theta,\Phi,\Psi) h_\times \label{eq:3.7}
	\end{equation}
where $F_+$ and $F_\times$ are the detector beam patterns given in
Thorne~\cite{Thorne_K:1987}.  In searching for continuous
gravitational waves from a particular direction, the Euler angles
become periodic function of sidereal time, thus resulting in an
amplitude and phase modulation of the observed
signal~\cite{Thorne_K:1987,Bonazzola_S:1996,Livas_J:1987}.  For
observation times longer than one sidereal day, the amplitude
modulation effectively averages the reception over all values of right
ascension, and over a range of declination which depends one the
precise position of the pulsar.  In particular, the effect of this
process is to allow detection of continuous waves from any direction,
but at the cost of reducing the measured strain (see
Fig.~\ref{figure3}).

\subsection{Parameter space}\label{parameter-space}

To facilitate later discussion it is useful to parameterize the
gravitational waveform by a vector $\bl=(\lambda^0,\vec\lambda)$ such
that
	\begin{equation}
	(\lambda^0,\lambda^1,\ldots\lambda^{s+2}) =
	(f_0,n_x,n_y,f_1,\ldots,f_s) \label{eq:3.8}
	\; .
	\end{equation}
Here $s$ is the maximum number of spindown parameters included in the
frequency model determined by Eq.~(\ref{eq:2.1}).  These vectors
span an $s+3$ dimensional space on which $\lambda^\alpha$ can be
thought of as coordinates.  (Note that $n_z^2 = 1-n_x^2-n_y^2$ is not
an independent parameter.)  In particular we denote the observed phase
of the gravitational waveform by
	\begin{equation}
	\phi(t;\bl) = 2 \pi \int^t\!\! dt'\; 
	f_{\rm \protect\scriptsize gw}(t') \; ,
	\label{eq:3.9}
	\end{equation}
where $f_{\rm \protect\scriptsize gw}(t')$ is given by Eq.~(\ref{eq:3.6}).

Initial interferometers in LIGO should have reasonable sensitivity to
gravitational waves with frequencies
	\begin{equation}
	f \ge 40\mbox{\rm Hz}\; , \label{eq:3.10}
	\end{equation}
while advanced interferometers are expected to have improved
sensitiviy down to
	\begin{equation}
	f \ge 10\mbox{\rm Hz} \; . \label{eq:3.11}
	\end{equation}
Moreover, theoretical constraints suggest that pulsars with spin
periods significantly smaller than one millisecond are unlikely.
This helps to constrain the highest frequency that one may wish to
consider in an all sky search to be about $2\mbox{\rm kHz}$.  According
to the discussion in section~\ref{pulsars}, the spindown parameters satisfy
	\begin{equation}
	-\tau_{\rm \protect\scriptsize min}^{-k} \le f_k \le \tau_{\rm
\protect\scriptsize min}^{-k} \; ,
	\label{eq:3.12}
	\end{equation}
where $\tau_{\rm \protect\scriptsize min}$ is the minimum spindown age of a
pulsar to be
searched for.  Finally, $n_x$ and $n_y$ are restricted by the relation
	\begin{equation}
	n_x^2+n_y^2 \le 1 \; . \label{eq:3.13}
	\end{equation}

\section{Data analysis technique}\label{data-analysis-technique}

Radio astronomers are familiar with searching for nearly periodic
sources in the output of their
detectors~\cite{Anderson_S:1993,Middleditch_J:1975}. The technique
employed by them is directly applicable to the problem at
hand~\cite{Livas_J:1987,Jones_G:1995}.

\begin{figure}
\psfig{file=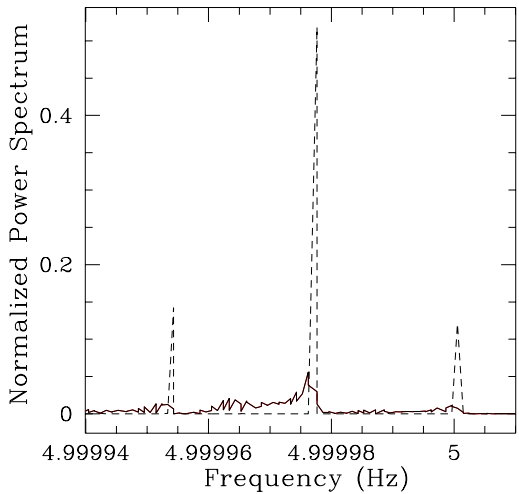,width=8cm,bbllx=0in,bblly=2.2in,%
bburx=8.5in,bbury=9.5in}
\caption{\label{figure3} Power spectra for two signals, each with
gravitational wave frequency $5\mbox{\rm Hz}$, computed using approximately
$10$
days worth of data; they are normalised with respect to the maximum
power achieved if the source was directly above an interferometer
which remained stationary during the entire observation.  The signal
was assumed to come from declination $0^\circ$ and right ascension
$90^\circ$; in fact the amplitude modulation is only sensitive to
changes in declination. The detector latitude was chosen to coincide
with LIGO detector in Hanford Washington.  The solid line corresponds
to a Doppler and amplitude modulated gravitational wave signal. The
dashed line is the same signal but with the Doppler modulation removed
by stretching.  The (unreasonably) low frequency was chosen for
illustrative purposes, so that both curves could appear on the same
scale.  For realistic gravitational wave frequencies ($\sim 500\mbox{\rm Hz}$)
the Doppler modulated signal would be further reduced by roughly two
orders of magnitude.}
\end{figure}

In the detector frame the gravitational wave signal can be
written as
	\begin{equation}
	h(t;\bl)= {\mbox{\rm Re}}\left[ {\cal A} e^{-i\phi(t;\bl)}
	\right] \label{eq:4.1}
	\end{equation}
where ${\cal A}=(h_{0+}+i h_{0\times})$,  $h_{0+}=F_+ (1+\cos^2 i)
h_0$ and $h_{0\times}= 2 F_\times ( \cos i ) \, h_0$.  The orbital phase
$\phi (t;\bl)$ is given by Eqs.~(\ref{eq:3.9}) and (\ref{eq:3.6}).  Introducing
a canonical time
	\begin{equation}
	t_b [t;\vec\lambda] = \frac{\phi(t;\bl)}{2 \pi f_0} \; ,
			\label{eq:4.2}
	\end{equation}
the above signal becomes monochromatic as a function of $t_b$. (The
presence of the amplitude modulation complicates the following
analysis without changing the conclusions significantly; therefore, we
treat $\cal A$ as constant in this and the next
section\footnote{Amplitude modulation can be viewed as the convolution
of the exactly periodic signal with some complicated window function.
Thus, in reality, the power spectrum of a stretched signal will not be
a monochromatic spike at a single frequency, but will be split into
several discrete, narrow spikes spread over a bandwidth $\delta f
\simeq 10^{-4}\mbox{\rm Hz}$.  After a preliminary detection, the
amplitude modulation spikes would provide a discriminant against false
signals~\cite{Livas_J:1987}.}.)  Figure~\ref{figure3} shows the
normalised power spectrum computed from the signal as a function of $t$
in Eq.~(\ref{eq:4.1}) (with $f_k\equiv 0$), compared with the spectrum from
the signal as a function of $t_b$.  It is clear that the maximum power
per frequency bin is significantly reduced when frequency modulation
is not accounted for.

Radio astronomers refer to this technique of introducing a canonical
time coordinate as {\em stretching} the data.  Since interferometer
output will be sampled at approximately $16\mbox{\rm kHz}$, in a practical
search for pulsars up to $2\mbox{\rm kHz}$ gravitational wave frequency, the
stretching can probably be achieved by resampling the
data stream appropriately.  This method, which is called {\em
stroboscopic sampling} by Schutz~\cite{Schutz_B:1991}, has the
benefit of keeping the computational overhead introduced by the
stretching process to a minimum.  We will return to this issue in a
later publication.

Now, a search of the detector output, $o(t)$, for gravitational waves
from a known source is straightforward.  One assumes specific
parameter values $\vec\xi$ in the waveform~(\ref{eq:4.1}), computes the
demodulated time function $t_b[t;\vec\xi\,]$ using Eq.~(\ref{eq:4.2}) and
stretches the detector output accordingly, thus
	\begin{equation}
	o_b(t_b[t;\vec\xi\,]) = o(t) \; .
	\label{eq:4.3}
	\end{equation}
If the assumed parameters $\vec\xi$ are not too much different from
the actual parameters $\vec\lambda$ of the signal, the stretched data
will consist of a nearly monochromatic signal.  One then takes the
Fourier transform with respect to $t_b$,
	\begin{equation}
	\tilde o(f;\vec\xi\,) = \frac{1}{\sqrt{ T_b^{\mbox{\rm
	\protect\scriptsize obs}}}} 
	\int_0^{T_b^{\mbox{\rm \protect\scriptsize obs}}}
	e^{2\pi i f t_b} \; o_b(t_b) \; dt_b \; .
	\end{equation}
Here $T_b^{\mbox{\rm \protect\scriptsize obs}}$ is length of the observation
measured using $t_b$.  The power spectrum is then searched for excess
power.  (The threshold is set by demanding some overall statistical
significance for a detection; see section~\ref{all-sky}.)  Notice that
the gravitational wave frequency, $\lambda^0=f_0$, is treated somewhat
differently than the other parameters; the Fourier transform searches
over all possible values in a single pass.  Given a sampled data set
containing $N$ points, the entire process, from original data through
to the power spectrum, requires of order $3N \log_2 N$ floating point
operations (to first approximation).

If all the parameters are not known accurately in advance, it will be
necessary to search over some of the remaining parameters
$\vec\lambda$; a separate demodulation and FFT must be performed for
each independent point in parameter space that one wishes to
search.   There are many possible refinements on this strategy which
could reduce the computational cost of a search by circumventing
certain stages of the procedure described here.   We mention some of
them in section~\ref{future},  however,  we focus attention on this
baseline strategy in this paper.

One more issue that arises in the discussion of stretching is how it
effects the noise in the detector.  Throughout this paper we assume
that the noise in the detector is a stationary, gaussian process;
however, when we stretch the output data stream the noise is no longer
strictly stationary unless it is perfectly white.  Real detectors will
have coloured noise, with correlations between points sampled at
different times.  Stretching the data modifies these correlations in a
time dependent manner.  In our case this is a very small effect,
having a characteristic timescale of several hours, and besides this
the noise in real detectors may be intrinsically non-stationary on
similar timescales due to instrumental effects.  Correcting pulsar
searches for such non-stationarity is an important problem, but one
that we do not address here.  We simply assume that $S_n(f)$, the
power spectral density of the noise, can be estimated on short
timescales and used in the conventional way for signal to noise
estimates.  Moreover, the effects of stretching on noise are only a
consideration when the noise is not white; since stretching affects
the power spectrum only within bands $\sim 10^{-1}\mbox{\rm Hz}$ wide, the
detector spectrum can usually be taken as white, unless we are near a
strong feature in the noise spectrum.  The precise nature of these
effects is being explored by Tinto~\cite{Tinto_M:1997}.

\section{Parameter space metric}\label{metric}

In general, neither the position of the pulsar nor its intrinsic
spindown may be known in advance of detection.  Therefore, the above
process, or some variant on it, must be repeated for many different
vectors $\vec\xi$ until the entire parameter space has been explored.
How finely must one sample these parameters in order to minimise the
risk of missing a signal?  A similar question arises in the context of
searching for signals from coalescing compact binaries using matched
filtering; Owen~\cite{Owen_B:1996} has introduced a general framework
to provide an answer in that case.  We adapt his method to the problem
at hand by defining a distance function on our parameter space; the
square of distance between two points in parameter space is
proportional to the fractional loss in signal power due to imprecise
matching of parameters.  The number of discrete points which must be
sampled can then be determined from the proper volume of the parameter
space with respect to this metric.

\subsection{Mismatch}\label{mismatch}

The one-sided power spectral density (PSD) of the detector output,
stretched with parameters $\vec\xi$, is
	\begin{equation}
	P_o (f) = 2\, | \tilde o(f;\vec\xi) |^2 \; .  \label{eq:5.1}
	\end{equation}
Now, suppose a detector output consists of a
signal with parameters $\bl$, and stationary, gaussian noise $n(t)$
such that
	\begin{equation}
	o(t) = h(t;\bl) + n(t) \; . \label{eq:5.2}
	\end{equation}
Thus, the expected PSD of the detector output, once again stretched
with parameters $\vec\xi$, is
	\begin{equation}
	E[\, P_o(f) \,] = 2\, | \tilde h(f;\bl,\Delta\vec\lambda)%
	|^2 + S_n(f) \; , \label{eq:5.3}
	\end{equation}
where $\Delta\vec\lambda = \vec\xi -\vec\lambda$, and $S_n(f)$ is the
one-sided power spectral density of the detector noise.  (As discussed
at the end of the previous section, we ignore the small effects of
stretching on the noise.)  The notation $\tilde
h(f;\bl,\Delta\vec\lambda)$ indicates the Fourier transform of a
signal, with parameters $\bl$, with respect to a time coordinate
$t_b[t;\vec\lambda+\Delta\vec\lambda]$.  We define the {\em
mismatch\/} $m(\bl,\Delta\bl)$ to be the fractional reduction in
signal power caused by stretching the data with the wrong parameters,
{\em and\/} by sampling the spectrum at the wrong frequency;
specifically,

\begin{equation} m(\bl,\Delta\bl) = 1 - \frac{|\tilde
h(f;\bl,\Delta\vec\lambda)|^2}{|\tilde h(f_0;\bl,0)|^2} \;
. \label{eq:5.4} \end{equation}

Remember that $\bl =
(\lambda^0=f_0,\vec\lambda)$.

\begin{figure}
\psfig{file=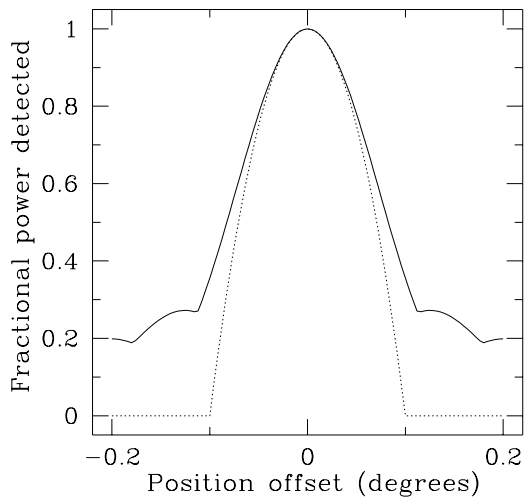,width=8cm,bbllx=0in,bblly=2.2in,%
bburx=8.5in,bbury=9.5in}
\caption{\label{figure4}
Fractional reduction in measured signal power caused by demodulating
by mismatched parameters (in this case, an error in the assumed
declination of the source).  The solid curve is the true power ratio,
the dotted is that given by the quadratic approximation of the metric.
Note that the widths of the curves agree well down to 70\% power
reduction ($m\sim 0.7$), beyond which the metric approximation
significantly underestimates the range of parameters permitted for a
specified power loss.  The curves are computed for a sky position of
$0^\circ$ right ascension, $45^\circ$ declination,  and no spindown.}
\end{figure}

In the present circumstance,  it is sufficient to consider a complex signal
	\begin{equation}
	h(t;\bl) = {\cal A} e^{-2\pi i f_0 t_b[t;\vec\lambda]}\; ,
		\label{eq:5.5}
	\end{equation}
where the amplitude $\cal A$ is constant.   The function
$t_b[t;\vec\lambda]$,  computed using Eqs.~(\ref{eq:4.2}), (\ref{eq:3.9}) and
(\ref{eq:3.6}), is explicitly written as
	\begin{equation}
	t_b[t;\vec\lambda] = \int_0^{t} dt'
	\{(1+{\textstyle \sum} f_k t'^k)(1 + \vec v\cdot\hat{n}/c)\}
	\; ,
	\label{eq:5.13}
	\end{equation}
Now,  the Fourier transform $\tilde h(f;\bl,\Delta\vec\lambda)$ is
	\begin{equation}
	\tilde{h}(f;\bl,\Delta\vec\lambda) = \frac{{\cal A}}{\sqrt{
	T_b^{\mbox{\rm \protect\scriptsize obs}}}}
	\int_0^{T_b^{\mbox{\rm \protect\scriptsize obs}}} \! d\hat{t}_b\; e^{i
	\Phi[t;\bl,\Delta\vec\lambda]} \; , \label{eq:5.6}
	\end{equation}
where
	\begin{equation}
	\frac{\Phi[t;\bl,\Delta\vec\lambda]}{2\pi} = \Delta
	\lambda^0 \hat{t}_b + f_0(
	{t}_b[t;\vec\lambda+\Delta\vec\lambda]
	- t_b[t;\vec\lambda]) \label{eq:5.7}
	\end{equation}
and $\Delta \lambda^0= f-f_0$.  Here,  $t$ should be interpreted as a
function of $\hat{t}_b$ defined implicity by $\hat{t}_b =
t_b[t;\vec\lambda+\Delta\vec\lambda]$.   Using
Eqs.~(\ref{eq:5.13})-(\ref{eq:5.7})
it is easy to show that $m(\bl,\Delta\bl)$ has a local
minimum of zero when $\Delta\bl \equiv 0$;
	\begin{eqnarray}
	m(\bl,\Delta\bl) |_{\Delta\bl = 0} &=& 0 \; ,
		\label{eq:5.8}\\
	\left. \partial_{\Delta\lambda^\alpha}m (\bl,\Delta\bl)
	\right|_{\Delta\bl = 0} &=& 0 \; . \label{eq:5.9}
	\end{eqnarray}
Thus, an expansion of the mismatch in powers of $\Delta\bl$ is
	\begin{equation}
	m(\bl,\Delta\bl) = \sum_{\alpha,\beta} g_{\alpha\beta}(\bl)
	\Delta\lambda^\alpha  \Delta\lambda^\beta + {\cal O}(\Delta
	\bl^3) \; , \label{eq:5.10}
	\end{equation}
where
	\begin{equation}
	g_{\alpha\beta} = {\textstyle\frac{1}{2}}\sum_{\alpha,\beta} \left.
	\partial_{\Delta\lambda^\alpha}
	\partial_{\Delta\lambda^\beta} m(\bl,\Delta\bl)
	\right|_{\Delta\bl = 0} \; . \label{eq:5.11}
	\end{equation}
In this way the mismatch defines a local distance function on the
signal parameter space, and, for small separations $\Delta\bl$,
$g_{\alpha\beta}$ is the metric of that distance function.  Note that
the metric formulation (\ref{eq:5.10}) will generally {\em overestimate\/}
the mismatch for large separations, as demonstrated in
Figure~\ref{figure4}.

Calculations using this formalism are considerably simplified by
partially evaluating the right hand side of Eq.~(\ref{eq:5.11}).  The form of
the signal~(\ref{eq:5.5}) allows us to write
	\begin{equation}
	g_{\alpha\beta}(\bl) =
	\left< \partial_{\Delta\lambda^\alpha} \Phi
	\partial_{\Delta\lambda^\beta} \Phi \right> - \left<
	\partial_{\Delta\lambda^\alpha} \Phi \right> \left<
	\partial_{\Delta\lambda^\beta} \Phi \right> \; ,
	\label{eq:5.12}
	\end{equation}
where $\Phi$ is given by Eq.~(\ref{eq:5.7}),  and where we use the notation
	\begin{equation}
	\left< \ldots \right> = \left. \frac{1}{T_b^{\mbox{\rm
	\protect\scriptsize obs}}}
	\int_0^{T_b^{\mbox{\rm \protect\scriptsize obs}}}\!\! (\ldots )
	dt_b \right|_{\Delta\bl=0} \; .
	\label{eq:5.14}
	\end{equation}

\subsection{Metric and number of patches}\label{number-patches}

Up until now, we have treated the frequency of the signal as one of
the parameters, $\lambda_0$, which must be matched.  In our search
technique, stretching and Fourier transforming the data yields an
entire power spectrum, automatically sampling all possible
frequencies.  We would really like to know the number of times that this
combination of procedures must be performed in a search.  This
requires knowledge of the mismatch $m(\bl,\Delta\bl)$ as a function of
$\Delta\vec\lambda$, having already maximized the power (i.e.\
minimized $m$) over $\lambda^0$.  The result is the mismatch projected
onto the $(s+2)$-parameter subspace:
	\begin{equation}
	\mu = \min_{\lambda_0} m(\bl,\Delta\bl) = \sum_{ij} \gamma_{ij}
		\Delta\lambda^i \Delta\lambda^j \label{eq:5.15}
	\end{equation}
where
	\begin{equation}
	\gamma_{ij} = g_{ij} - \frac{g_{0i} g_{0j}}{g_{00}}\; ,
		\label{eq:5.16}
	\end{equation}
and $i=1,\ldots,s+2$.  We will generally refer to $\mu$ as the
{\em projected mismatch}.

Technically, $\gamma_{ij}$ should be computed from $g_{\alpha\beta}$
evaluated at the specific value of $\lambda_0$ at which the minimum
projected mismatch occurred.  However, since this number is unknown in advance
of detection, we evaluate $\gamma_{ij}$ for the largest frequency in
the search space.  In this way we never underestimate the projected mismatch.

In a search, the parameter space will be sampled at a lattice of
points, chosen so that no location in the space has $\mu$ (given by
Eq.~(\ref{eq:5.15}) greater than some
$\mu_{\rm \protect\scriptsize max}$ away from one of the points.  This is
equivalent to
tiling the parameter space with {\em patches} of maximum
extent $\mu_{\rm \protect\scriptsize max}^{1/2}$.  The number of points we must
sample at is
therefore
	\begin{equation}
	N_p = \frac{\int_{\cal P} \sqrt{\det||\gamma_{ij}||} \,
		d^{s+2}\vec\lambda} {V_{\rm \protect\scriptsize patch}}
		\; , \label{eq:5.17}
	\end{equation}
where $V_{\rm \protect\scriptsize patch}$ is the proper volume of a single
patch, and $s+2$
is the reduced dimensionality of the parameter space $\cal P$
(excluding $\lambda_0$).

Optimally,  one should use some form of spherical closest packing to
cover the space with the fewest patches.  Our solution uses hexagonal
packing in two of the dimensions and cubic packing in all the others;
in this way the volume of a single patch is
	\begin{equation}
	V_{\rm \protect\scriptsize patch} = \frac{3\sqrt{3}}{4} 
		\left(\frac{4 \mu_{\rm
\protect\scriptsize max}}
		{s+2}\right)^{(s+2)/2}
	\end{equation}

\section{Depth of an all sky search}\label{all-sky}

We are finally in a position to estimate the depth of a search for
periodic sources using LIGO.  The detector participates in two
principal motions which cause significant Doppler modulations of the
observed signal: daily rotation, and revolution of the Earth about the
Sun.  The latter is actually a complex superposition of an elliptical
Keplerian orbit with a smaller orbit about the earth-moon barycenter,
and is further perturbed by interactions with other planets.  For now,
however, we use a simplified model which treats both rotation and
revolution as circular motions about separate axes inclined at an
angle $\epsilon=23^\circ 27'$ to each other.  Although a
simplification, this does remove any spurious symmetries from the
model; thus, an actual search using the precise ephemeris of the earth
in its demodulations should give comparable results.  In this model,
then, we write the velocity of the detector in a frame which is
inertial to the solar system barycenter but initially comoving with
the earth:
	\begin{eqnarray}
	\vec v &=& -(\Omega R_d \sin \Omega t - \Omega_A R_A
		\sin \Omega_A t) \vec x  \nonumber\\
	&&+ (\Omega R_d \cos \Omega t - \Omega_A R_A \cos
		\epsilon [\cos \Omega_A t - 1]) \vec y \label{eq:6.1}\\
	&& -\Omega_A R_A \sin\epsilon [\cos \Omega_A t - 1]) \vec z
	\nonumber
	\end{eqnarray}
where $R_d= 6.371\times10^8 (\cos l)$cm, $l$ is the latitude of the
detector, and $R_A=1.496\times10^{13}$cm is the distance from the
earth to the sun.  The angular velocities are
$\Omega=2\pi/(86\,400\mbox{s})$ and $\Omega_A =
2\pi/(3.155674\times10^7\mbox{s})$.  Our coordinate system measures
$\vec x$ towards the vernal equinox and $\vec z$ towards the north
celestial pole, and we arbitrarily choose to measure time starting at
noon on the vernal equinox.

The number of spindown parameters $f_k$ which must be included to
account for all intrinsic frequency drift depends to a large extent on
the type of pulsar one wishes to search for.  We determined this
number on a case by case basis, including all parameters which lead to
a significant increase in the number of parameter space patches.
Equivalently, the following geometric picture suggests a simple
criterion for deciding when there is one spindown parameter too many
included in the signal parametrization.  Let $\lambda^L$ be the last,
`questionable' spindown parameter $f_s$ (so $L=s+2$).  With respect to
the natural metric $\gamma_{ij}$ on parameter space, the unit-normal
to surfaces of constant $\lambda^L$ is just $\gamma^{iL}/
(\gamma^{LL})^{1/2}$, where $\gamma^{ij}$ is the inverse of
$\gamma_{ij}$.  The spindown parameter $\lambda^L$ is unnecessary if
the proper thickness of the parameter space in this normal direction
is less than half the proper grid spacing; that is, if
$2 \tau_{\rm \protect\scriptsize min}^{-L+2}/(\gamma^{LL})^{1/2} <
\sqrt{\mu_{\rm \protect\scriptsize max}/L}$.
In practice, one has included more spindown parameters than necessary
if and only if $\gamma^{LL} > 4 L \tau_{\rm \protect\scriptsize min}^{-2L+2} /
\mu_{\rm \protect\scriptsize max}$.

\subsection{Patch number versus observation time}

\begin{figure}[t]
\vbox{
\psfig{file=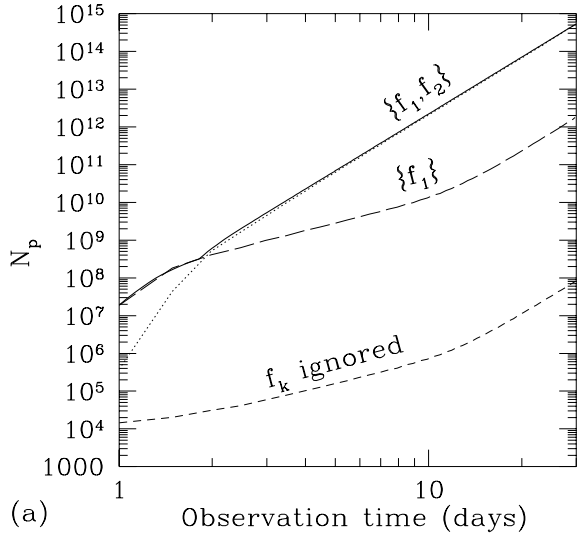,width=8.5cm,bbllx=0in,bblly=2.2in,%
bburx=8.5in,bbury=10in}
\psfig{file=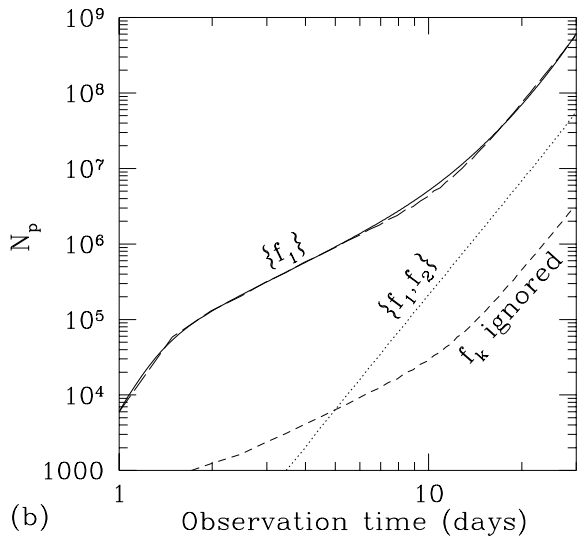,width=8.5cm,bbllx=0in,bblly=2.2in,%
bburx=8.5in,bbury=10in}
}
\caption{ \label{figure5}
Number of independent points in parameter space as a function of total
observation time, using a maximum projected mismatch
$\mu_{\rm \protect\scriptsize max}=0.3$.  The
parameter ranges chosen were: (a) maximum gravitational wave frequency
$1\,000\mbox{\rm Hz}$, minimum spindown age $\tau_{\rm \protect\scriptsize
min}=40\mbox{\rm Yrs}$
(hypothetical young pulsars); (b) maximum gravitational wave frequency
$200\mbox{\rm Hz}$, minimum spindown age $\tau_{\rm \protect\scriptsize
min}=10^3\mbox{\rm Yrs}$ (observed,
slow pulsars).  The short-dashed curve represents the total number of
patches ignoring all $f_k$.  The long-dashed curve is the number of
patches including only $f_1$ in the search.  The dotted line is the
number of patches including both $f_1$ and $f_2$.  Also shown is the
empirical fit given in the text; it was normalised by the results in
shown in (a).  In some regimes, searching over an additional spindown
parameter would seem to reduce the number of patches; however, this
actually only indicates regions where the parameter space extends less
than one full patch width in the additional dimension.  In such
regimes one must properly discard the extra parameter from the search,
forcing one to choose always the higher of the curves.}
\end{figure}

It is extremely difficult to obtain a closed form expression for the
metric, let alone its determinant.  Therefore, we present results for
two concrete scenarios which suggest themselves based on the
discussion in section~\ref{pulsars}: (i) hypothetical sources with $f_0 \le
1000\mbox{\rm Hz}$, and spindown ages greater than $\tau = 40\mbox{\rm Yrs}$;
incidentally, this also includes the majority of known, millisecond
pulsars; and (ii) slower sources ($f_0 \le 200\mbox{\rm Hz}$) having spindown
ages
in excess of $\tau = 1000\mbox{\rm Yrs}$.  The number of parameter space points
which must be searched is plotted as a function of total observation
time in Fig.~\ref{figure5}.  The numbers are normalised by a maximum
projected mismatch $\mu_{\rm \protect\scriptsize max}=0.3$.

In considering an optimal choice of observation time, it is useful to
construct an empirical fit to $N_p(t_{\rm \protect\scriptsize obs},\mu_{\rm
\protect\scriptsize max})$.  Notice
first that all the parameters $\Delta\vec\lambda$ in $\Phi$, given by
Eq.~(\ref{eq:5.7}), appear multiplied by the gravitational wave frequency
$f_0$;  thus, $N_p \propto (f_0)^{s+2}$.  Furthermore, provided the
determinant of the metric is only weakly dependent on the values of
the $f_k$ one may also extract a factor of $\tau^{-s(s+2)/2}$; our
investigations suggest the validity of this approach.
In this way we arrive at the expression
	\begin{equation}
	N_p \simeq \max_{s\in\{0,1\ldots\}} \left[
		{\cal N}_s
		F_s(t)  \right] \; , \label{eq:6.2}
	\end{equation}
where
	\begin{eqnarray}
	&&{\cal N}_s =
	\left( \frac{f_0}{1\mbox{\rm kHz}} \right)^{s+2} \left(
	\frac{40\mbox{\rm Yrs}}{\tau} \right)^{s(s+1)/2}
	\left( \frac{0.3}{\mu_{\rm \protect\scriptsize max}} \right)^{(s+2)/2}
	\label{eq:6.3}\\
	&&F_0(t_{\rm \protect\scriptsize obs}) = 6.9\times10^3\; T^2 + 3.0\;
		T^5\label{eq:6.4}\\
	&&F_1(t_{\rm \protect\scriptsize obs}) = \frac{1.9\times10^8\; T^8 +
5.0\times10^4
		\; T^{11}}{4.7 + T^6}
		\label{eq:6.5} \\
	&&F_2(t_{\rm \protect\scriptsize obs}) = \frac{2.2\times10^7\;
		T^{14}}{56.0 + T^9}\; ,\label{eq:6.6}
	\end{eqnarray}
and $T$ is the observation time measured in days.  These formulae are
normalised using only the data corresponding to Fig~\ref{figure5}(a), and
subsequently compared with computed values for several frequencies and
spindown ages $\tau$.  The analytic fit is in good agreement with the
computed results for a variety of parameters; however, the fits
generally break down for observation times less than one day.
We stress that more spindown parameters may become important for
observation times longer than 30 days.

Schutz~\cite{Schutz_B:1991} has previously estimated the number of points
which must be searched in the abscence of spindown corrections;   he
argued that this number scaled as $T^4$ for observation times longer
than about a day.   The difference between his previous estimate and
the expression in Eq.~(\ref{eq:6.4}),  which shows that the number of points
increases as $T^5$,  derives from an asymmetery between declination
and right ascension which was not accounted for in his argument.

\begin{figure}[t]
\psfig{file=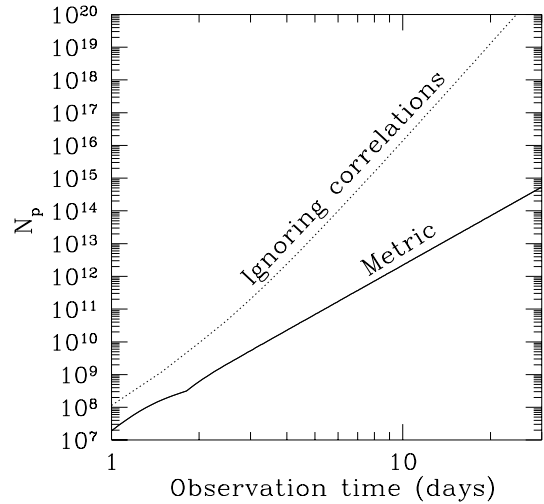,width=8cm,bbllx=0in,bblly=2.2in,%
bburx=8.5in,bbury=9.5in}
\caption{\label{figure5a} The total number of parameter-space points
needed to search for pulsars having gravitational wave frequency up to
$1\mbox{\rm kHz}$, and spindown age greater than $\tau=40\mbox{\rm Yrs}$.  The
solid line
is the number computed using the metric and properly accounting for
correlations between various terms in the frequency evolution.  The
dotted line is the same number computed directly by assuming the
points must lie on the grid of coordinates used to parameterise the
signal.  The benefits of using the metric to optimally place the
points to be searched in parameter space is clear.
}
\end{figure}

The benefit of the metric formulation is that it accounts for the
significant correlations which exist between the intrinsic spindown
and the earth-motion-induced Doppler modulations by using points which
lie on the principal axes of the ellipsoids described by
Eq.~(\ref{eq:5.15}).   Replacing the invariant volume integral in
Eq.~(\ref{eq:5.17}) by
	\begin{equation}
	\int_{\cal P} \sqrt{{\textstyle\prod_i} \gamma_{ii}}\;
	d^{s+2}\vec\lambda
	\; ,
	\end{equation}
gives the number of points required for a search if, instead,  one
chooses them to lie on the $\{n_x,n_y,f_1,f_2,...\}$ coordinate grid.
Figure~\ref{figure5a} shows the total number of points computed using
this method compared to the results obtained using the invariant
volume integral.  For sufficiently long integration times the
difference can be several orders of magnitude.

\subsection{Computational Requirements}

The number of real samples of the interferometer output for an
observation lasting $t_{\rm \protect\scriptsize obs}$ seconds, and sampled at a
frequency
$2f_{\rm \protect\scriptsize max}$, where $f_{\rm \protect\scriptsize max}$ is
the maximum gravitational wave
frequency being searched for, is
	\begin{equation}
	N = 2f_{\rm \protect\scriptsize max} 
		t_{\rm \protect\scriptsize obs} \; .
\label{eq:6.8}
	\end{equation}
For each $\vec\lambda$ that is used to stretch the detector output, a
search then requires an FFT, calculation of the power, and some
thresholding test for excess power.  Assuming that the stretching and
thresholding require negligible computations compared to performing
the FFT and computing the power, the total number of floating point
operations for a search is
	\begin{equation}
	N_{\rm \protect\scriptsize op} = 6 f_{\rm \protect\scriptsize max} 
	t_{\rm\protect\scriptsize obs} 
	N_p [\log_2 (2f_{\rm \protect\scriptsize max}
		t_{\rm \protect\scriptsize obs}) + 1/2]\; , \label{eq:6.9}
	\end{equation}
where $N_p$ is given by~(\ref{eq:6.2})-(\ref{eq:6.6}).  The additive $1/2$
inside
the square brackets accounts for the three floating point operations
per frequency bin which is required to compute the power from the Fourier
transform.

A guideline for a feasible, long-term, search strategy is that data
reduction should proceed at a rate comparable to data acquisition.
Thus,  the total computing power required for data
reduction, in floating point operations per second (flops), is
	\begin{equation}
	P = \frac{N_{\rm \protect\scriptsize op}}
	{t_{\rm \protect\scriptsize obs}} =
		6 f_{\rm \protect\scriptsize max} 
		N_p(t_{\rm \protect\scriptsize
		obs},\mu_{\rm \protect\scriptsize max})
		[\log_2 (2f_{\rm \protect\scriptsize max}%
		t_{\rm \protect\scriptsize obs}) +
		1/2] \; . \label{eq:6.10}
	\end{equation}
For a prescribed maximum projected mismatch $\mu_{\rm \protect\scriptsize
max}$, and maximum
available computing power $P_{\rm \protect\scriptsize max}$ this expression
determines the
maximum allowed coherent integration time.  Alternatively, given the
computing power available for data reduction, $P_{\rm \protect\scriptsize
max}$, it
provides an implicit relation between $\mu_{\rm \protect\scriptsize max}$ and
the
integration time.

The idea now is to choose $\mu_{\rm \protect\scriptsize max}$ and $t_{\rm
\protect\scriptsize obs}$ so that we
maximize the sensitivity of the search.  In order to do this we must
first obtain a threshold, above which we consider excess power to
indicate the presence of a signal.

As discussed in section~\ref{data-analysis-technique}, we assume that
the noise in the detector is a stationary, gaussian random process
with zero mean and PSD $S_n(f)$.  In the abscence of a signal, the
power $P_o(f) = 2 |\tilde n(f)|^2$ is exponentially distributed with
probability density function
\begin{equation}
\frac{e^{-P_o(f)/S_n(f)}}{S_n(f)} \; . \label{eq:6.11}
\end{equation}
We assume that there is independent noise in each of
$f_{\rm \protect\scriptsize max}t_{\rm \protect\scriptsize obs}$ frequency bins
for a given demodulated power
spectrum.  In general the noise spectra obtained from neighbouring
parameter space points will not be statistically independent; however,
one may expect that the correlations will be small when the mismatch
between the points approaches unity.  Therefore we approximate the
number of statistically independent noise spectra in our search to be
$N_p(t_{\rm \protect\scriptsize obs},\mu_{\rm \protect\scriptsize max}=0.3)$.
In order that a detection have
overall statistical significance $\alpha$, we must set our detection
threshold so there is less than $1-\alpha$ probability of {\em any\/}
noise event exceeding that threshold.  For a detection to occur the
power in the demodulated detector output must satisfy
\begin{equation}
\frac{P_o(f)}{S_n(f)} > \frac{\rho_c}{S_n(f)} =
\ln\left[ \frac{f_{\rm \protect\scriptsize max}t_{\rm \protect\scriptsize obs}
N_p
(t_{\rm \protect\scriptsize obs},\mu_{\rm \protect\scriptsize max}=0.3)}
{1-\alpha}\right] \; , \label{eq:6.12}
\end{equation}
where $P_o(f)$ was defined in Eq.~(\ref{eq:5.1}), and $\rho_c$ is the
threshold power.

In other words, if the power at a given frequency exceeds
$\rho_c$ we can infer that a signal is present; the expected power in
the signal is then $\rho_c-S_n$.  Thus, the minimum characteristic
amplitude we can expect to detect is
	\begin{equation}
	h_{\rm \protect\scriptsize th} = \sqrt{\frac{(\rho_c/S_n-1)S_n(f)}
	{<F_+^2(\Theta,\Phi,\Psi)> (1-<\mu>)\,
		t_{\rm \protect\scriptsize obs}}}
		\; , \label{eq:6.13}
	\end{equation}
where $<F_+^2(\Theta,\Phi,\Psi)>$ is the square of the detector
response averaged over all possible source positions and wave
polarizations.  $<\mu>$ is the expected mismatch for a source whose
signal parameters $\vec\lambda$ lie within a given patch, assuming
that all parameter values in that patch are equally likely.  We note
that the characteristic detector sensitivities $h_{\rm \protect\scriptsize
3/yr}$ in
Fig.~\ref{figure1} are obtained from this expression by setting
$t_{\rm \protect\scriptsize obs}=10^7$~seconds, $<\mu>=0$, and $f_{\rm
\protect\scriptsize max}t_{\rm \protect\scriptsize obs}N_p =
1$ in the expression for $\rho_c$; this agrees with Eq.~(\ref{eq:1.2}).

\begin{figure}
\vbox{
\psfig{file=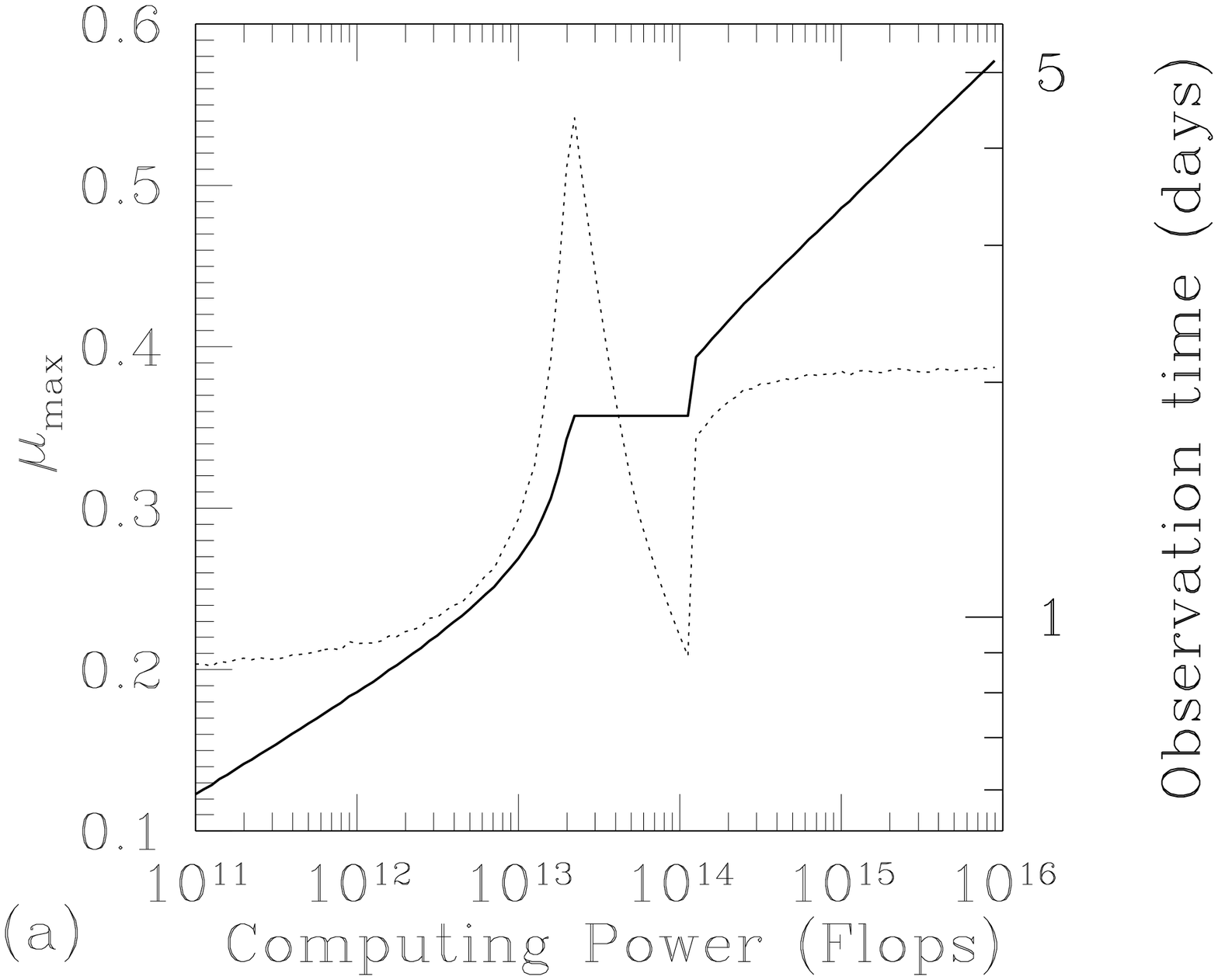,width=9cm,bbllx=0in,%
bblly=1.8in,bburx=8.5in,bbury=8.5in}
\psfig{file=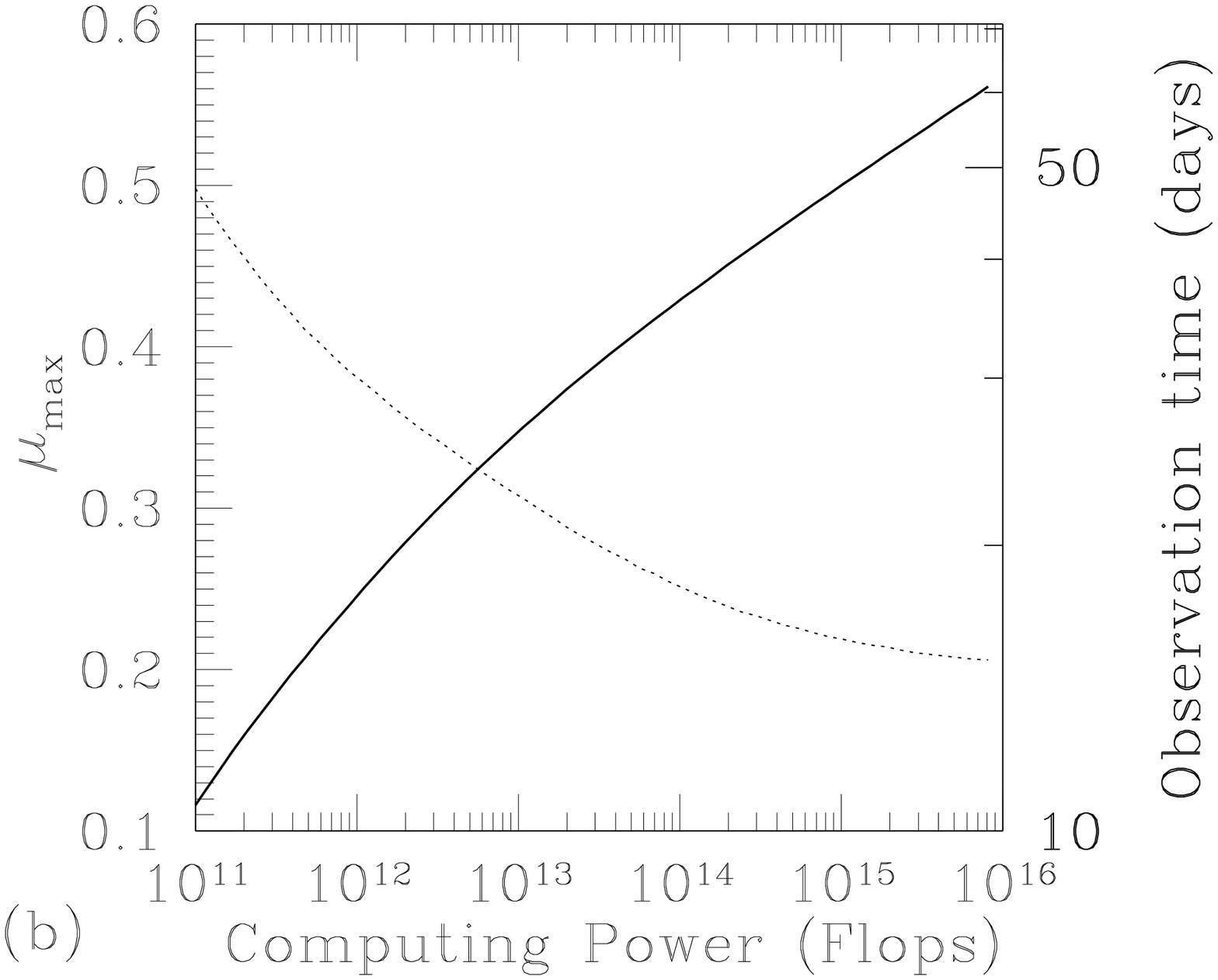,width=9cm,bbllx=0in,%
bblly=1.8in,bburx=8.5in,bbury=8.5in}
}
\caption{\label{figure6} The optimum observation time (thick solid
line), and maximal projected mismatch (dotted line) as functions of
available computational power.  Both graphs assume a threshold which
gives an overall statistical significance of $99\%$ to any detection
(although the results should be insensitive to the precise value).
Each of the graphs corresponds to: (a) the situation encountered when
searching for periodic sources having gravitational wave frequencies
up to $1000\mbox{\rm Hz}$, with minimum spindown ages $\tau_{\rm
\protect\scriptsize min}=40\mbox{\rm Yrs}$.
(b) The equivalent results for gravitational wave frequencies up to
$200\mbox{\rm Hz}$, with minimum spindown ages $\tau_{\rm \protect\scriptsize
min}=10^3\mbox{\rm Yrs}$.  The
transition region seen in figure~(a) is due to the fact that a longer
integration time would require searching over an additional spindown
parameter, as seen in Fig.~\protect\ref{figure5}.  In this region it is more
efficient, as
one adds computational power, to lower mismatch thresholds, rather
than searching over the additional parameter.}
\end{figure}

The optimal search strategy is to choose those values of $t_{\rm
\protect\scriptsize obs}$
and $\mu_{\rm \protect\scriptsize max}$ which, for some specified computational
power
$P_{\rm \protect\scriptsize max}$ and detection confidence $\alpha$, maximize
our {\it
sensitivity} $\Theta$ which is defined by
	\begin{equation}
	\Theta(t_{\rm \protect\scriptsize obs},\mu_{\rm 
	\protect\scriptsize max})
\equiv \frac{1}{h_{\rm \protect\scriptsize th}}
		\propto \sqrt{\frac{[1-\frac{s+2}{s+4}\mu_{\rm 
	\protect\scriptsize max}]T}
				{\rho_c/S_n-1}} \; \label{eq:6.14},
	\end{equation}
where $\rho_c/S_n$ is given by Eq.~(\ref{eq:6.2}).  Assuming an overall
statistical significance of $\alpha=0.99$, we have computed the
optimal observation time $t_{\rm \protect\scriptsize obs}$ and optimal maximum
mismatch
$\mu_{\rm \protect\scriptsize max}$, as functions of computing power, for the
two searches
considered in the previous subsection.  The results are shown in
Fig.~\ref{figure6}.

\section{ Computational Requirements for a Directed Search} \label{directed}

  In sections \ref{metric} and \ref{all-sky} we examined the
computational requirements of an all-sky pulsar search.  In this
section we examine the computational requirements for a directed
pulsar search, by which we mean a search where the position is known
but the pulsar frequency and spin-down parameters are unknown.
Obvious targets in this category are SN1987A, nearby supernova
remnants that do not contain known radio pulsars, and the center of
our galaxy. Such searches will clearly be among the first performed
once the new generation of gravitational wave detectors begin to come
on line.

  Our treatment of directed pulsar searches closely parallels that of
of the all-sky search, so we can be brief. Since the source position
$(n_x, n_y)$ is known, we can simply remove the Earth's motion from
the data. Below we imagine that the signal has already been transformed
to the solar system barycenter.
Then the unknown parameters describing the pulsar waveform are
	\begin{equation}
	(\lambda^0,\lambda^1,\ldots\lambda^s) =
	(f_0,f_1,\ldots,f_s) \label{eq:7.1}
	\; .
	\end{equation}
\noindent where the $f_i$ are the same as defined in Eq.~(\ref{eq:2.1}) and
$s$ is just the number of spindown parameters included in the
frequency model. We again calculate the metrics $g_{ij}$ and
$\gamma_{ij}$ using Eqs.~(\ref{eq:5.12}) and (\ref{eq:5.16}) respectively, and
then calculate $N_p$ using (\ref{eq:5.17}) (except the integral is now over
s-dimensional parameter space).   Assuming hexagonal packing in two
dimensions and cubic packing in the others,  the size of each patch is
$V_{\rm \protect\scriptsize patch} = (3\sqrt{3}/4) (4 \mu_{\rm
\protect\scriptsize max}/s)^{s/2}$.  (Except for $s=1$, where
$V_{\rm \protect\scriptsize patch} =
2 \mu_{\rm \protect\scriptsize max}^{1/2}$.)  We arrive at the expression
	\begin{equation}
	N_p \simeq \max_{s\in\{1,2\ldots\}} \left[
		{\cal N}_s
		G_s(t)  \right] \; , \label{eq:7.2}
	\end{equation}
where
	\begin{eqnarray}
	&&{\cal N}_s =
	\left( \frac{f_0}{1\mbox{\rm kHz}} \right)^s \left(
	\frac{40\mbox{\rm Yrs}}{\tau} \right)^{s(s+1)/2}
	\left( \frac{0.3}{\mu_{\rm \protect\scriptsize max}} \right)^{s/2}
	\label{eq:7.2a}\\
	&&G_1(t_{\rm \protect\scriptsize obs}) = 
		1.5\times10^3\; T^2 \label{eq:7.3}\\
	&&G_2(t_{\rm \protect\scriptsize obs}) = 
		6.97\times10^{1}\; T^5 \label{eq:7.4}\\
	&&G_3(t_{\rm \protect\scriptsize obs}) = 2.89\times10^{-4}\; T^9
,\label{eq:7.5}
	\end{eqnarray}
\noindent where $T$ is the observation time measured in days.
Comparing these results with Eqs.~(\ref{eq:6.2})--(\ref{eq:6.6}), we see that
for
our fiducial parameter values ($f_0 = 1{\rm kHz}$, $\tau_{\rm
\protect\scriptsize min} = 40
{\rm Yrs}$, $\mu_{\rm \protect\scriptsize max} = 0.3$) and observation times
$T$ of order a
week, $N_p$ is $\sim 10^5$ times larger for an all-sky search than for
a directed search. Another way of putting this is: after using one's
freedom to adjust the frequency and spin-down parameters in optimizing
the fit, only $\sim 10^5$ distinguishable patches
on the sky remain.  Equivalently, a single directed search can cover an area
of $\sim 10^{-4}$ steradians. Thus $\sim 1000$ week-long, directed
searches would be sufficient to cover the galactic center region.

\begin{figure}
\vbox{
\psfig{file=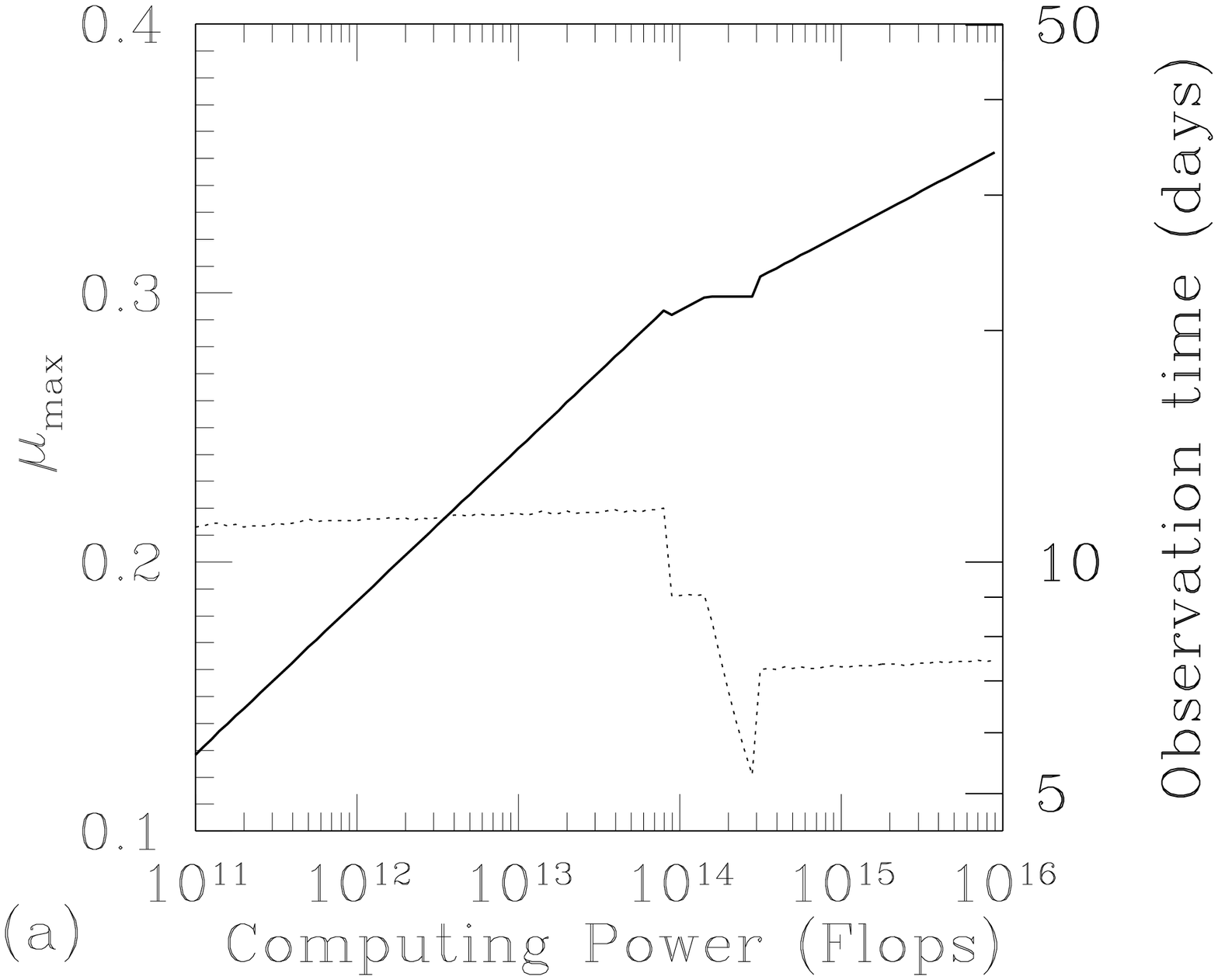,width=9cm,bbllx=0in,%
bblly=1.8in,bburx=8.5in,bbury=8.5in}
\psfig{file=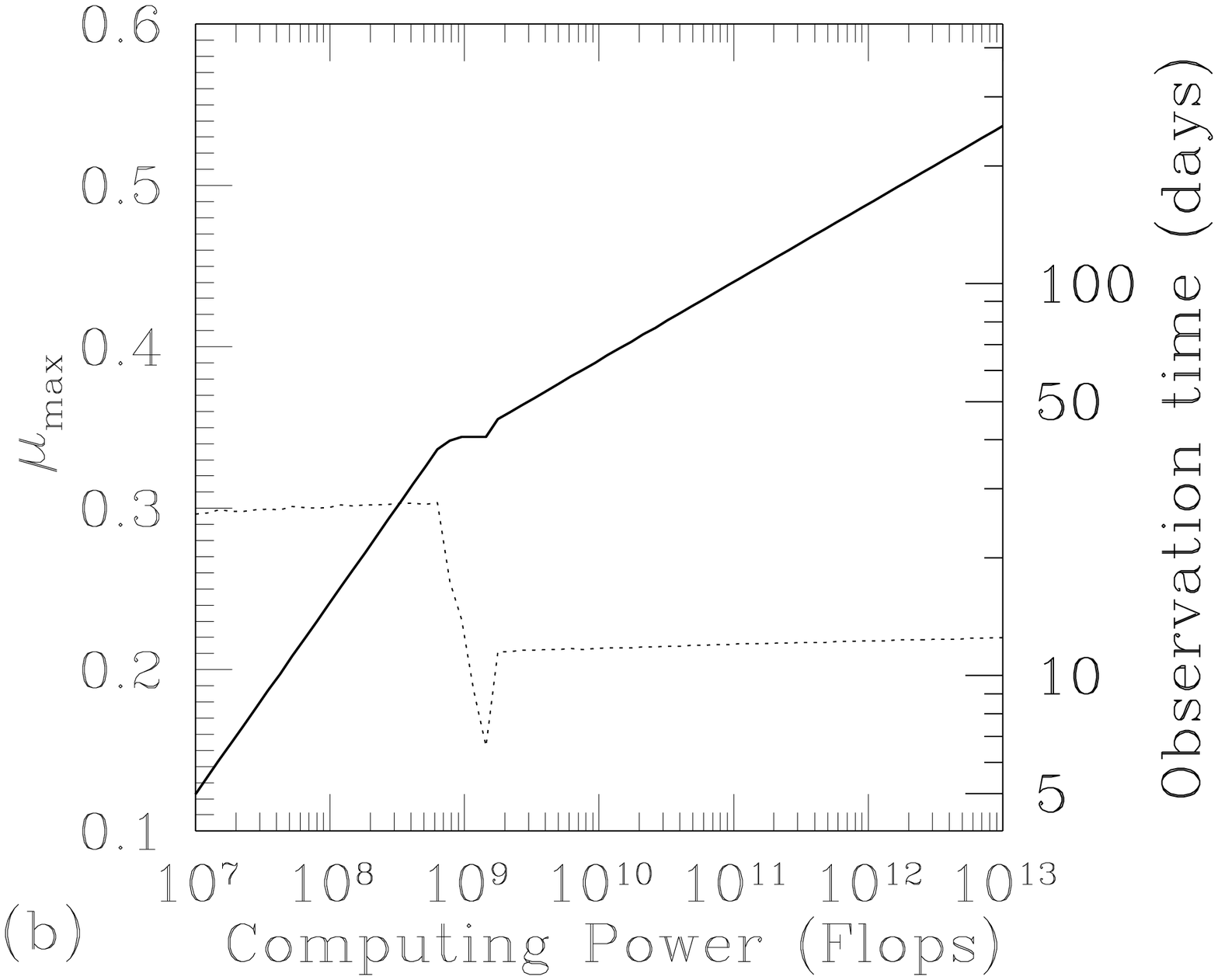,width=9cm,bbllx=0in,%
bblly=1.8in,bburx=8.5in,bbury=8.5in}
}
\caption{\label{figure7} The optimum observation time (thick solid
line), and maximal projected mismatch (dotted line) as functions of
available computational power for directed searches.  Both graphs
assume a threshold which gives an overall statistical significance of
$99\%$ to any detection (although the results are insensitive to the
precise value).  Each of the graphs corresponds to: (a) the situation
encountered when searching for periodic sources having gravitational
wave frequencies up to $1000\mbox{\rm Hz}$, with minimum spindown ages
$\tau_{\rm \protect\scriptsize min}=40\mbox{\rm Yrs}$.  (b) The equivalent
results for gravitational
wave frequencies up to $200\mbox{\rm Hz}$, with minimum spindown ages
$\tau_{\rm \protect\scriptsize min}=10^3\mbox{\rm Yrs}$.  The transition
regions, where the optimum
observation time does not increase, are due to the fact that a longer
integration time would require searching over an additional spindown
parameter.}
\end{figure}

We can calculate the optimal $\mu_{\rm \protect\scriptsize max}$ and $t_{\rm
\protect\scriptsize obs}$ as a
function of computing power for a directed search in the same way as
we did for the all-sky directed search.  (Except the factor ${{s+2}\over{s+4}}$
in Eq.~\ref{eq:6.14} becomes ${{s}\over{s+2}}$ for the directed-search case.)
The results are shown in
Fig.~\ref{figure7}, for our two fiducial types of pulsar.  We see that
knowing the source position in advance increases $t_{\rm \protect\scriptsize
obs}$ by only
a factor of $\sim 10$, for 1~Tflops computing power.  The resulting
gains in sensitivity can be seen in Fig.~\ref{figure1a}.

\section{Future directions}\label{future}

Searching for unknown sources of continuous gravitational waves using
LIGO, or other interferometers, will be an immense computational task.
In this paper we have presented our current understanding of the
problem.  By applying techniques from differential geometry we have
estimated the number of independent points in the parameter space
which must be considered in all-sky and directed searches for sources
which spin down on timescales short enough to produce observable
effects; these numbers were used to compute the maximum achievable
sensitivity for a coherent search (see Fig.~\ref{figure1a}).
Furthermore, the metric formulation can be used to optimally place the
parameter space points which must be sampled in a search.

Our analysis takes no account of bottlenecks in the analysis process
due to data input/output and inter-processor communication.  These are
important issues which may impose further constraints on the maximum
observation time; however, it seems premature to address such problems
until we know the hardware that will be used to conduct searches for
continuous waves.

Unfortunately, Fig.~\ref{figure6} shows that it will be impossible to
search, in one step, $10^7$ seconds worth of data over all sky
positions.  However it is also unnecessary.  We foresee implementing a
hierarchical search strategy, in which a long data stream is searched
in two (or more) stages, trading off sensitivity in the first stage
for reduced computational requirements.  Having determined a number of
potential signals in the first stage---presumably at a threshold level
which allows many false alarms due to random noise---these candidate
events would be followed up in the second stage, using longer
integration times.  The longer integration times would be possible
because the search would only have to be performed over much smaller
regions of the parameter space, in the neighbourhoods of the candidate
signal parameters.  In this way, one can achieve a greater sensitivity
than a coherent search using the same computational resources.

Clearly one can imagine many different implementations of this rough
strategy, and we have not yet determined the optimal one.
Nevertheless, we have considered the simple example where the data is
searched in two stages.  Candidate signals from an all-sky search of a
short stretch of data [$T^{(1)}$ seconds long] are followed up using
longer Fourier transforms to achieve greater sensitivity.  One can
estimate $T^{(1)}$ using Fig.~\ref{figure6} and an assumption
that roughly half of the total computing budget is used on the first
stage; this turns out to be a valid assumption.  A simple argument
along these lines goes as follows.  Consider a search for `young,
fast' pulsars that begins by coherently analyzing stretches of data
that are all $\sim 1$ day long (possible with $\sim 4\times 10^{12}$
flops, by Fig.~\ref{figure6}).  Imagine that in the second stage of
the search one follows up all templates such that $P_0(f,\vec\lambda)
> 4.6 S_n(f)$, by seeing whether templates with roughly the same
parameter values are exceeding this threshhold every day.  (Here
$P_0(f,\vec\lambda)$ is the power of the stretched data at frequency
$f$, for stretch $\vec\lambda$. This threshhold implies that one is
following up only one out of every hundred templates.)  It seems
likely that this second stage will not be more computationally
intensive than the first. To exceed this threshold, a pulsar must have
$h_c \agt 12 h_{\rm \protect\scriptsize 3/yr}$. This is factor of roughly $3$
better than
if one restricted oneself to coherent searches considered above, but
is a factor of $3$ worse than the sensitivity one could achieve with
unlimited computing power.

A refinement of this strategy would be one in which the first pass
consists of several incoherently-added power spectra.  That is, one
slices the data into $N$ sequential subsets, performs a full search
(as described in this paper) for each subset, and adds up the power
spectra of the resulting searches for each of the parameter sets.
This technique has been used to good effect by radio astronomers
searching for pulsars~\cite{Anderson_S:1993}.  Since the addition of
power spectra is incoherent, there is a loss of signal-to-noise ratio
in the final summed power spectrum of $1/\sqrt{N}$ in relation to a
full coherent search over the whole timescale.  However, the
computational savings involved allow one to search stretches of data
which are much longer overall.  For some optimal choice of $N$, this
will result in higher sensitivities when one follows up candidate
detections using coherent searches.  D.~Nicholson (private
communication) has estimated that a $1\mbox{\rm Tflops}$ computer could perform
such a search of $10^7$s of data, over all sky positions but ignoring
pulsar spindowns.  A subsequent paper will present a concrete analysis
of this and other hierarchical scenarios~\cite{Creighton_T:1996}.

\acknowledgements

PRB and TC wish to thank Kip Thorne for suggesting that we work on
this problem; we are extremely grateful to him for his encouragement
and suggestions.  PRB and TC also thank Bruce Allen, Kent Blackburn,
Jolien Creighton, Sam Finn, Scott Hughes, Ben Owen, Massimo Tinto, and
Alan Wiseman for useful discussions.  This work was supported in part
by NSF grant PHY-9424337.  PRB is supported by a PMA prize fellowship
at Caltech.  BFS wishes to thank Gareth Jones, Andrez Krolak, Andrew
Lyne and David Nicolson for useful conversations.  CC thanks Lee
Lindblom for helpful discussions; CC's work was supported in part by
NSF grant PHY-9507686 and by a grant from the Alfred P. Sloan
Foundation.


\begin{thebibliography}{10}

\bibitem{Abramovici_A:1992}
A. Abramovici {\it et~al.}, Science {\bf 256},  325  (1992).

\bibitem{Bradaschia_C:1990}
C. \mbox{Bradaschia~{\sl et al.}}, Nucl. Inst. A. {\bf 289},  518  (1990).

\bibitem{Blanchet_L:1996}
L. Blanchet, B.~R. Iyer, C.~M. Will, and A.~G. Wiseman, Class. Quant. {\bf 13},
   575  (1996).

\bibitem{Apostolatos_T:1996}
T.~A. Apostolatos, Phys. Rev. D. {\bf 54},  2421  (1996).

\bibitem{Owen_B:1996}
B. Owen, Phys. Rev. D. {\bf 53},  6749  (1996).

\bibitem{Balasubramanian_R:1996}
R. Balasubramanian, B.~S. Sathyaprakash, and S.~V. Dhurandhar, Phys. Rev. D.
  {\bf 53},  3033  (1996).

\bibitem{Thorne_K:1987}
K.~S. Thorne,  in {\em Three hundred years of gravitation}, edited by S.~W.
  Hawking and W. Israel (Cambridge University Press, Cambridge, 1987), Chap.~9,
  pp.\ 330--458.

\bibitem{Flanagan_E:1993}
E.~E. Flanagan, Phys. Rev. D. {\bf 48},  2389  (1993).

\bibitem{Compton_K:1996}
K. Compton, Ph.D. thesis, University of Wales, Cardiff, 1996.

\bibitem{Nicholson_D:1996}
D. Nicholson {\it et~al.}, Physics Letters {\bf A 218},  175  (1996).

\bibitem{Allen_B:1996}
B. Allen,  in {\em Proceedings of the Les Houches School on Astrophysical
  Sources of Gravitational Waves}, edited by J.~A. Marck and J.~P. Lasota
  (Cambridge University Press, Cambridge, 1996).

\bibitem{Bonazzola_S:1996}
S. Bonazzola and E. Gourgoulhon, Astron. Astr. {\bf 312},  675  (1996).

\bibitem{Chandrasekhar_S:1970}
S. Chandrasekhar, Phys. Rev. Let. {\bf 24},  611  (1970).

\bibitem{Friedman_J:1978}
J.~L. Friedman and B.~F. Schutz, Ap. J. {\bf 222},  281  (1978).

\bibitem{Wagoner_R:1984}
R.~V. Wagoner, Astrophys. J. {\bf 278},  345  (1984).

\bibitem{Zimmermann_M:1979}
M. Zimmermann and E. Szedenits, Jr., Phys. Rev. D. {\bf 20},  351  (1979).

\bibitem{Blandford_R:1984}
R.~D. Blandford, unpublished  (1984).

\bibitem{Schutz_B:1991}
B.~F. Schutz,  in {\em The Detection of Gravitational Waves}, edited by D.~G.
  Blair (Cambridge University Press, Cambridge, 1991), Chap.~16, pp.\ 406--451.

\bibitem{Livas_J:1987}
J.~C. Livas, Ph.D. thesis, Massachusetts Institute of Technology, 1987.

\bibitem{Jones_G:1995}
G.~S. Jones, Ph.D. thesis, University of Whales, 1995.

\bibitem{Niebauer_T:1993}
T.~M. Niebauer {\it et~al.}, Phys. Rev. D {\bf 47},    (1993).

\bibitem{Shapiro_SL:1983}
S.~L. Shapiro and S.~A. Teukolsky, {\em Black Holes, White Dwarfs and Neutron
  Stars} (Wiley, New York, 1983), Chap.~10.

\bibitem{Lyne_A:1992}
A.~G. Lyne, Phil. Trans. R. Soc. Lond. A {\bf 341},  29  (1992).

\bibitem{Narayan_R:1990}
R. Narayan and J.~P. Ostriker, Astrophys. J. {\bf 352},  222  (1990).

\bibitem{Manchester_R:1992}
R.~N. Manchester, Phil. Trans. R. Soc. Lond. A {\bf 341},  3  (1992).

\bibitem{Kulkarni_S:1992}
S.~R. Kulkarni, Phil. Trans. R. Soc. Lond. A {\bf 341},  77  (1992).

\bibitem{Wolszczan_A:1989}
A. Wolszczan {\it et~al.}, Nature {\bf 337},  531  (1989).

\bibitem{Anderson_S:1993}
S.~B. Anderson, Ph.D. thesis, California Institute of Technology, Pasadena,
  California, 1993.

\bibitem{Zimmermann_M:1978}
M. Zimmermann, Nature {\bf 271},  524  (1978).

\bibitem{Zimmermann_M:1980}
M. Zimmermann, Phys. Rev. D. {\bf 21},  891  (1980).

\bibitem{Galtsov_D:1984}
D.~V. Gal'tsov, V.~P. Tsvetkov, and A.~N. Tsirulev, Sov. Phys. ---JETP {\bf
  59},  472  (1984).

\bibitem{Middleditch_J:1975}
J. Middleditch, Ph.D. thesis, University of California, Berkeley, 1975.

\bibitem{Tinto_M:1997}
M. Tinto, private communication  (1997).

\bibitem{Creighton_T:1996}
P.~R. Brady and T. Creighton, paper in preparation  (1996).

\end{thebibliography}
\end{document}